\documentclass[preprint,1p]{elsarticle}
\usepackage{amsmath}
\usepackage{amssymb}
\usepackage{amsthm}
\usepackage{mathtools}
\usepackage{bm}
\usepackage{hyperref}
\usepackage[capitalise]{cleveref}
\usepackage{caption}
\usepackage{subcaption}
\usepackage{booktabs}
\usepackage{multirow}
\usepackage{siunitx}
\usepackage{interval}
\usepackage{enumitem}
\usepackage{lineno}
\usepackage{xcolor}

\journal{Signal Processing}
\hypersetup{colorlinks=true}
\graphicspath{{./}}

\newcommand\scalemath[2]{\scalebox{#1}{\mbox{\ensuremath{\displaystyle #2}}}}

\newcommand{\NORM}[2]{{\left\lVert#1\right\rVert}_{#2}}

\newcommand*{\DB}{\ensuremath{\unit{\deci\bel}}}

\DeclareMathOperator{\diag}{diag}
\DeclareMathOperator*{\argmin}{arg\,min}

\newtheorem{property}{Property}
\newtheorem{proposition}{Proposition}
\newtheorem{corollary}{Corollary}
\newtheorem{remark}{Remark}
\allowdisplaybreaks{}

\makeatletter
\def\ps@pprintTitle{%
    \let\@oddhead\@empty
    \let\@evenhead\@empty
    \def\@oddfoot{\footnotesize
        {\textit{Preprint accepted to \@journal{},} \textsc{doi}: \href{https://doi.org/10.1016/j.sigpro.2025.109944}{10.1016/j.sigpro.2025.109944} \hfill \textit{February 9, 2025}}%
    \let\@evenfoot\@oddfoot
    }
}
\makeatother

\begin{document}
\begin{frontmatter}
	\title{Joint Time-Vertex Fractional Fourier Transform}
	\author[aff1,aff2]{Tuna Alika\c sifo\u glu}
	\author[aff3]{B\"unyamin Kartal}
	\author[aff4]{Eray \"Ozg\"unay}
	\author[aff1,aff2]{Aykut Ko\c c\corref{cor1}}
	\cortext[cor1]{Corresponding author}
	\ead{aykut.koc@bilkent.edu.tr}

	\address[aff1]{Department of Electrical and Electronics Engineering, Bilkent University, Ankara, T\"urkiye}
	\address[aff2]{UMRAM, Bilkent University, Ankara, T\"urkiye}
	\address[aff3]{WINS Lab, Massachusetts Institute of Technology (MIT), Cambridge, MA, USA}
	\address[aff4]{Politecnico di Milano, Milano, Italy}

	\begin{abstract}
		Graph signal processing (GSP) facilitates the analysis of high-dimensional data on non-Euclidean domains by utilizing graph signals defined on graph vertices. In addition to static data, each vertex can provide continuous time-series signals, transforming graph signals into time-series signals on each vertex. The joint time-vertex Fourier transform (JFT) framework offers spectral analysis capabilities to analyze these joint time-vertex signals. Analogous to the fractional Fourier transform (FRT) extending the ordinary Fourier transform (FT), we introduce the joint time-vertex fractional Fourier transform (JFRT) as a generalization of JFT\@. The JFRT enables fractional analysis for joint time-vertex processing by extending Fourier analysis to fractional orders in both temporal and vertex domains. We theoretically demonstrate that JFRT generalizes JFT and maintains properties such as index additivity, reversibility, reduction to identity, and unitarity for specific graph topologies. Additionally, we derive Tikhonov regularization-based denoising in the JFRT domain, ensuring robust and well-behaved solutions. Comprehensive numerical experiments on synthetic and real-world datasets highlight the effectiveness of JFRT in denoising and clustering tasks that outperform state-of-the-art approaches.
	\end{abstract}

\end{frontmatter}

\section{Introduction}\label{sec:intro}
With rapidly increasing technology, there has been a substantial increase in data stored and processed worldwide. A significant portion of this data is collected on networks that inherently have irregular underlying structures. Due to this irregularity, these networks are represented as graphs, where the data residing on them are represented as time-varying graph signals. The graph signal processing (GSP) framework and the graph neural networks (GNNs) have become prevalent approaches for such problems~\cite{sandryhaila13discretegsp,shuman13emerging,sandryhaila13filtersicassp,ortega18gspsurvey,ortega22gsptextbook,yan23spectralgfrt,sheng24samplingtv,zach24gsinter,wei24gfrt,xuan22avgnet,bronstein17gnnreview,wu21gnnreview}.

Graphs are essential to represent data collected from irregular and complex networks and, therefore, are extensively studied and used~\cite{feizi20spectralalign, klickstein19gengraphs,gong21distmulagent,kim14allrandomcom,watts98collectivedynamic,barabasi99scalingrandomnet}. Data can be represented as signals residing on the vertices of a graph. Hence, GSP became one of the upcoming tools for processing network data. Classical signal processing methods have been generalized to the GSP domain, including sampling and approximation~\cite{chen15sampling,anis14sampling,narang13interpolation, zhu12apprxgraphsignals}, filtering~\cite{liu19filter,onuki16denoising,gavili17shiftoperator,ribeiro18gsp,wang18gspfrtsampling, hua19graphfilter,isufi24graphfilterml}, Fourier transformation and its duality~\cite{cheung20deepGSP,leus21dualgraphshift, kartal21vm}, and frequency analysis~\cite{sandryhaila14freq,grassi18timevertex}. The graph Fourier transform (GFT), which is the generalization of the Fourier transform (FT) for graph signals, along with other GSP techniques, gave rise to numerous applications including smoothing and denoising~\cite{onuki16denoising,shuman11chebyshev,loukas13reshape}, segmentation~\cite{loukas14globaltrends}, graph signal reconstruction~\cite{giraldo22sobolev,qui17graphsigrecons}, classification~\cite{berger17graphsignalrec,bayram18lidar}, clustering~\cite{belkin02laplacianeigenmaps,maretic20laplacian}, low-rank extraction~\cite{shahid16fastpca}, estimation~\cite{perraudin17stationary,marques17stationary,kroizer22bayesianestgraph}, non-stationary analysis~\cite{hammond11graphwavelets,coifman06diffusionwavelet}, semi-supervised learning~\cite{belkin04semisup}, multiscale decompositions~\cite{hammond11graphwavelets,zheng16multiscaledec,narang12waveletgraph}, stationary process processing~\cite{perraudin17stationary,marques17stationary}, signal prediction~\cite{kwak21traffic}, inference~\cite{wai22gspinference}, graph learning~\cite{bayram20mask,egilmez19learning, yang17learning}, intrusion detection~\cite{sadreazami18distributed}. A wide variety of applications of GNNs on several network architectures also exist, such as machine learning applications~\cite{dong20gspforml}, deep learning structures~\cite{cheung20deepGSP}, classification~\cite{xuan22avgnet,ye21robust,aras24grte,aras24textrgnn}, graph convolutional networks (GCNs)~\cite{liu21selfgconv,such17robustfiltgcnn,zhang19gcnreview,alikasifoglu24vispool}, denoising networks~\cite{chen21unrollingnetwork}, infection analysis~\cite{hosseinalipour19infectiongsp}, spatiotemporal data applications~\cite{pan2021spatiotemporal,das21spatiotemporal}.

While vertices and edges are primarily interested in modeling graph data, the time-domain information in vertices is also intertwined and essential to modeling network data. Moreover, time-domain information is generally a natural extension. For example, while the distribution of weather stations in a region forms a graph, these stations also record daily and, in some cases, hourly measurements. Similarly, sensor networks are naturally interconnected and have a history of measurements. In other words, data defined on the vertices of graphs change over time and applications like the above make analysis of joint time-vertex signals necessary. This spurs a need for a framework to process the temporal graph data jointly with the vertex domain. The tools developed by the pioneering works of~\cite{grassi18timevertex,loukas16jft,perraudin17towardsstationary,loukas19stationarytimevertex} combine the temporal discrete signal processing and GSP\@. Referring to time-vertex signal processing, these essential techniques enable us to work with time-varying graph data by considering temporal and graph-domain information. As the primary joint Fourier analysis tool to obtain spectral expansions of time-varying graph signals, the joint time-vertex Fourier transform (JFT) has been developed~\cite{grassi18timevertex,loukas16jft}. JFT combines FT in the temporal domain and GFT in the vertex domain~\cite{grassi18timevertex,loukas16jft}. Time domain information in time-vertex signals may be deterministic~\cite{grassi18timevertex} or stationary processes~\cite{loukas19stationarytimevertex}. Autoregressive moving average (ARMA) and vector autoregressive moving average (VARMA) filter for processing of time-vertex graph signals is developed~\cite{isufi17armafilter,isufi19forecasting}. Time-vertex signal processing is deployed in several applications, including reconstruction of time-varying graph signals~\cite{qiu17timevar,mao19timevarreconst, giraldo20sobolevc19}, predicting the joint spectral temporal data~\cite{isufi19forecasting}, and predicting the evolution of stationary graph signals~\cite{loukas17predict}. There are also applications of semi-supervised learning and inpainting of joint time-vertex signals~\cite{perraudin17towardsstationary,loukas19stationarytimevertex,loukas17predict}. Spatio-temporal graph applications can also be seen as joint time-vertex signals, and there are transform and filtering considerations for such applications~\cite{pan2021spatiotemporal,das21spatiotemporal}.

On the other hand, we have the fractional Fourier transform (FRT) as a generalization of the classical FT that allows for intermediate transformations between time and frequency domains, parameterized by a transform order that dictates the fraction of the transformation. The $\alpha^{\text{th}}$ order FRT is defined as the $\alpha^{\text{th}}$ power of the ordinary FT~\cite{ozaktas93frt,mendlovic93mainFRT1,ozaktas01book,pei00DLCTdef,koc24trainable}. The FRT reduces to the FT and the identity operations when $\alpha$ = 1 and $\alpha$ = 0, respectively. The ordinary FT is a transformation between time (or space) signals into spectral signals, whereas FRT converts intermediate domains between time (or space) and frequency (spatial frequency). Thus, it can be viewed as a linear transformation corresponding to a $\frac{\alpha\pi}{2}$ degree rotation in the time-frequency plane. The FRT also has the property of additive indexes, meaning that the $\alpha^{\text{th}}$ order of $\beta^{\text{th}}$ order FRT is equal to the ${(\alpha+\beta)}^{\text{th}}$ order FRT\@.

The FRT is a fundamental transform with important applications in several fields, including signal processing, optics, and wave propagation~\cite{ozaktas01book}. The applications of FRT in signal processing include time-frequency analysis~\cite{mustard96timefreq}, filter design~\cite{kutay97frtfiltering,zalevsky96wiener}, image processing~\cite{lohmann93imagerotation,jindal14video}, video processing~\cite{jindal14video}, time-series processing~\cite{koc22frfttimeseries}, natural language processing~\cite{sahinuc22frfttransformer}, beamforming~\cite{ahmad19frtadaptive}, pattern recognition~\cite{mendlovic98frtpatternrecog}, phase retrieval~\cite{dong97frtphaseret}, optical information processing~\cite{ozaktas93frt}, sonar signal processing~\cite{jacob09frtsonar}, inverse synthetic-aperture radar (ISAR) imaging~\cite{zhao18frtradar} among numerous others. FRT provides extra degrees of freedom when transforming signals into intermediate time-frequency domains while keeping the ordinary FT as a special case. This feature of FRT gives flexibility in processing data and makes performance improvements possible, mostly without additional computational costs.

Similarly, extending FRT to the GSP domain can open up further developments, application areas, and performance increases. To this end, graph fractional Fourier transform (GFRT), which transforms graph signals into intermediate vertex-frequency or vertex-spectral domains, has been introduced~\cite{wang18gspfrtsampling,wang17gspfrt,yan20windowedgfrftcon,yan21windowfrftjou, wu20fracspecgraphwavelet,alikasifoglu24unified}. Furthermore, windowed fractional Fourier transform has been generalized to GSP~\cite{yan20windowedgfrftcon,yan21windowfrftjou} and sampling in fractional domains is studied~\cite{wang18gspfrtsampling}. Along with several joint time-vertex transforms~\cite{yan22multi,zhang24jlct,zhang24hilbert}, Wiener and optimal filtering have recently been studied in the GSP domain~\cite{ozturk21graphfilter,ge23optimal,alikasifoglu24wiener}, where the optimal filtering happens in intermediate domains.

In this work, we introduce the joint time-vertex fractional Fourier Transform (JFRT) to extend the recent theoretical studies of joint time-vertex signal processing. The JFRT allows the simultaneous analysis of signals in both graph fractional and time fractional domains. We show that JFRT has the properties of index additivity in both domains, reduction to the ordinary transformation when the orders are 1 in both domains, reduction to the identity operator when the orders are 0, commutativity, and reversibility. We also show that if ordinary GFT is unitary, so is the JFRT\@. JFRT reduces to the two-dimensional discrete fractional Fourier transform (DFRT) for specific graph topologies. To further develop the theory, we present the fractional joint time-vertex filtering and the underlying theoretical setting for Tikhonov regularization-based denoising by the proposed JFRT\@. To this end, we present some properties of fractional Laplacians and define the joint fractional Laplacian to derive the optimal filter coefficients.

The proposed JFRT, with the flexibility provided by its two fractional order parameters, can be utilized in joint time-vertex signal processing applications such as denoising, signal reconstruction, and graph-node classification since it enables the joint signal to be processed both in FRT and GFRT domains. This makes it possible to analyze joint time-vertex signals in a much broader class of transformations by extending the theory of ordinary JFT analysis. Since JFRT satisfies most of the underlying properties of the two-dimensional DFRT, it is also a good candidate for the generalization of multidimensional FRTs to GSP\@. On the other hand, the proposed JFRT also contributes to the well-established and rich literature on fractional Fourier analysis by extending the classical theory to GSP machinery. We can summarize the contributions of this work as follows:
\begin{itemize}
	\item We introduce the JFRT to simultaneously analyze time-vertex signals in both fractional time and fractional vertex domains.
	\item We derive Tikhonov regularization-based denoising in the JFRT domain to ensure that the solution is more robust and well-behaved, particularly in the presence of noisy data.
	\item We demonstrate the effectiveness of JFRT through comprehensive numerical experiments on synthetic and real-world datasets in filtering, denoising, and clustering tasks.
\end{itemize}

The rest of the manuscript is organized as follows. In Section~\ref{Section_background}, we provide preliminary information for GSP, GFT, JFT, and FRT\@. We introduce the JFRT in Section~\ref{sec_frac_jft} and provide its properties. In Section~\ref{tiknovossec}, we develop the Tikhonov regularization-based denoising in the JFRT domain. Section~\ref{sect_exp} shows the utility of JFRT through numerical experiments and examples in filtering, denoising, and clustering tasks. Section~\ref{conclusion} concludes the paper.

\section{Preliminaries}\label{Section_background}

\subsection{Notation}
Bold uppercase letters ($\mathbf{A}$) denote matrices, while lowercase capital letters ($\mathbf{x}$) denote vectors. Given a set $\mathcal{V}$, $|\mathcal{V}|$ denotes its cardinality. The ${(\cdot)}^*$, ${(\cdot)}^T$, ${(\cdot)}^H$ denote complex conjugate, transpose, and complex conjugate transpose (Hermitian) of their argument. If the argument of $\diag(\cdot)$ is an ordered set, it constructs a square matrix with elements of the ordered set as diagonal elements. If the argument is a matrix, it gives a column vector containing the matrix's diagonal entries. $\mathbf{I}_{n}$ denotes an identity matrix of size $n$, and we omit the subscript if the context is clear. For given matrices $\mathbf{A}$ and $\mathbf{B}$, $\mathbf{A} \otimes \mathbf{B}$ denote their Kronecker product, and if $\mathbf{A} \in \mathbb{C}^{m \times m}$, $\mathbf{B} \in \mathbb{C}^{n \times n}$ are square matrices, $\mathbf{A} \oplus \mathbf{B} = \mathbf{A} \otimes \mathbf{I}_n + \mathbf{I}_m \otimes \mathbf{B}$ denotes their Kronecker sum. For scalars $N$, $M$, the modulo $M$ of $N$ is denoted by ${(N)}_M$. For a vector $\mathbf{v}$, ${\lVert \mathbf{v} \rVert}_p$ denotes $p$-norm. For a matrix $\mathbf{A}$ of size $m$ by $n$, its Frobenius norm is defined as ${\Vert \mathbf{A} \Vert}_F = \sqrt{\sum_{m^{\prime} = 0}^{m-1} \sum_{n^{\prime} = 0}^{n-1} |\mathbf{A}_{m^{\prime}, n^{\prime}}|^2}$.

\subsection{Graph Signals and Graph Fourier Transform (GFT)}
Let $\mathcal{G} = \{\mathcal{V}, \mathcal{E}, \mathbf{A} \}$ be a graph with set of vertices denoted as $\mathcal{V} = \{v_0, v_1, \dots v_{N-1}\}$, where $|\mathcal{V}| = N \in \mathbb{Z}^+$, set of edges denoted as $\mathcal{E}$ and weighted adjacency matrix denoted as $\mathbf{A} \in \mathbb{C}^{N \times N}$. An edge $e = (m,n)$ is an element of $\mathcal{E}$  if $\mathbf{A}_{m,n} \neq 0$ where $\mathbf{A}_{m,n}$ denotes the element in the intersection of $m^{\text{th}}$ row and $n^{\text{th}}$ column, otherwise there is no connection from $v_m$ to $v_n$. A graph is undirected if $\mathbf{A}_{m,n} = \mathbf{A}_{n,m}$ for all $m,n \in  \{0,1, \dots, N-1\}$. A graph signal $\mathbf{x} \in \mathbb{C}^N$ is a mapping from $\mathcal{V}$ to $\mathbb{C}$~\cite{sandryhaila13discretegsp, sandryhaila13filtersicassp} such that $ \mathbf{x}: \mathcal{V} \rightarrow \mathbb{C}$ and $v_n \rightarrow \mathbf{x}_n$.

Several approaches define GFT~\cite{sandryhaila13discretegsp,shuman13emerging,shafipour19digraph,singh16directedlaplacian}. Among them, two main approaches stand out. The first one follows the algebraic signal processing framework that views the adjacency matrices as a shift operator and builds a GFT definition accordingly~\cite{sandryhaila13discretegsp}. The second one uses the graph Laplacian, which is defined for undirected graphs with non-negative weighted adjacency matrices~\cite{shuman13emerging}.

\subsubsection{Algebraic Signal Processing-Based Approach}
Let $\mathbf{A} = \mathbf{V} \mathbf{J}_{\mathbf{A}} \mathbf{V}^{-1}$ be the Jordan decomposition of $\mathbf{A}$.  Then, GFT and its inverse transform are defined as
\begin{equation}
	\mathbf{F}_G \mathbf{x} = \mathbf{V}^{-1}\mathbf{x} = \mathbf{\tilde{x}}\quad \text{ and }\quad
	\mathbf{F}_G^{-1} \mathbf{\tilde{x}} = \mathbf{V} \mathbf{\tilde{x}} = \mathbf{x},
\end{equation}
respectively, where $\mathbf{\tilde{x}}$ stands for the graph signal in the GFT domain with GFT matrix $\mathbf{F}_G \triangleq \mathbf{V}^{-1}$. This approach has the advantage that it can be used for any graph type while requiring the Jordan decomposition, which is computationally more expensive than the graph Laplacian-based approach for large graphs.

\subsubsection{Graph (Combinatorial) Laplacian-Based Approach}
\indent In the graph Laplacian-based approach, the adjacency matrix $\mathbf{A}$ takes non-negative real values and is assumed to be symmetric. The graph Laplacian can be given as $\mathbf{L} = \mathbf{D} -\mathbf{A}$ where $\mathbf{D}_{m,m} = \sum_{n = 0}^{N-1} \mathbf{A}_{m,n}$ is the diagonal degree matrix. Then, we have $\mathbf{L} = \mathbf{U} \Lambda \mathbf{U}^H$ as the diagonalization of $\mathbf{L}$. Since $\mathbf{L}$ is symmetric positive semi-definite, it is unitarily diagonalizable. Then, the GFT and its inverse transform are defined as the following~\cite{shuman13emerging}:
\begin{equation}
	\mathbf{F}_G\mathbf{x} = \mathbf{U}^{H}\mathbf{x} = \mathbf{\tilde{x}}
	\quad\text{ and }\quad
	\mathbf{F}^{-1} \mathbf{\tilde{x}} = \mathbf{U} \mathbf{\tilde{x}} = \mathbf{x},
\end{equation}
respectively, where $\mathbf{\tilde{x}}$ is the representation of graph signal $\mathbf{x}$ in the graph Fourier (spectral) domain with $\mathbf{F}_G \triangleq \mathbf{U}^H$. Although there are attempts for generalizations~\cite{singh16directedlaplacian}, the Laplacian-based approach is generally considered limited to undirected graphs. However, the Laplacian-based GFT has the advantage of providing a unitary transformation so that the Parseval's relation holds.

\begin{remark}
	As described in~\cite{ortega22gsptextbook}, these approaches are based on the spectral decomposition of any selected graph shift matrix $\mathbf{Z}$, where $\mathbf{Z}=\mathbf{A}$ for algebraic and $\mathbf{Z}=\mathbf{L}$ for Laplacian-based approaches.
\end{remark}

Hence, in this work, we consider any arbitrary invertible GFT matrix denoted by $\mathbf{F}_G$, which can be obtained for any graph with a given shift matrix through the Jordan decomposition of $\mathbf{Z}=\mathbf{V}_Z^{}\mathbf{J}_Z^{}\mathbf{V}_Z^{-1}$, where $\mathbf{F}_G\triangleq\mathbf{V}_Z^{-1}$.

\begin{remark}
	For completeness, we provide Jordan decomposition also to include non-diagonalizable graph shift matrices $\mathbf{Z}$. For diagonalizable matrices, the Jordan decomposition reduces to the eigendecomposition $\mathbf{Z}=\mathbf{V}_Z^{}\mathbf{\Lambda}_Z^{}\mathbf{V}_Z^{-1}$. Finally, for Hermitian matrices, $\mathbf{Z}=\mathbf{Z}^H$, $\mathbf{Z}=\mathbf{V}_Z^{}\mathbf{\Lambda}_Z^{}\mathbf{V}_Z^H$, with $\mathbf{F}_G\triangleq\mathbf{V}_Z^H$.
\end{remark}

\subsubsection{Graph Frequency Ordering}\label{sec:graph_freq_ordering}
Eigenvector ordering of the selected shift matrix, i.e., column ordering of \(\mathbf{F}_G\), consequently graph frequency ordering is unique up to a permutation. Different frequency ordering and consequently different \(\mathbf{F}_G\) definitions are possible according to the selected shift matrix $\mathbf{Z}$. As in~\cite{shuman13emerging}, for the undirected Laplacian-based GFT, all eigenvalues are real and non-negative, so the graph frequencies are ordered as the ascending eigenvalues of the Laplacian. However, it does not apply to the directed adjacency-based GFT, so \emph{total variation on graphs} (\(TV_G\)) approach has been proposed in~\cite{sandryhaila14freq}, which orders frequencies in descending \(TV_G(\mathbf{v}) = {\lVert \mathbf{v} - {|\lambda_{\max}|}^{-1}\mathbf{A}\mathbf{v}\rVert}_1\) values, for eigenvectors \(\mathbf{v}\) of \(\mathbf{A}\), where \(\lambda_{\max}\) is the largest magnitude eigenvalue of \(\mathbf{A}\). We use suitable ordering for the graph signal processing tasks in this work.

\subsection{Fractional Fourier Transform (FRT)}
The straightforward \emph{linear integral form} definition of continous $\alpha^{\text{th}}$ order FRT, $\mathcal{F}^\alpha$, is provided in~\cref{eq:frt:cont} for $f(u)\in\mathcal{L}^2(\mathbb{C})$, $\alpha \in \mathbb{R}$, $k\in\mathbb{Z}$. It is not the most intuitive way to understand the FRT, but it is the most general form. More details of FRT can be found in~\cite{ozaktas93frt,mendlovic93mainFRT1,ozaktas01book}.
\begin{align}\label{eq:frt:cont}
	\mathcal{F}^{\alpha} f(u) & = \int_{-\infty}^{\infty} f(u^\prime) K_{\alpha}(u,u^\prime) du^\prime, \nonumber \\
	A_\theta                  & = \sqrt{1 - j\cot\theta} \quad\text{ and }\quad \theta = \frac{\alpha\pi}{2},     \\
	K_{\alpha}(u,u^\prime)    & =
	\begin{cases}
		\delta(u-u^\prime),                                                               & \alpha=4k         \\
		\delta(u+u^\prime),                                                               & \alpha=4k+2       \\
		A_{\theta} e^{j\pi(u^2\cot\theta-2uu^\prime\csc\theta + {u^\prime}^2\cot\theta)}, & \text{otherwise.} \\
	\end{cases} \nonumber
\end{align}

The discretization of the FRT is done by the discrete fractional Fourier transform (DFRT)~\cite{candan00dfrt}, and each entry of the DFRT matrix is given as follows:
\begin{equation}
	\mathbf{F}^\alpha[m,n] = \sum_{k\in\mathcal{K}} \mathbf{u}_k[m]e^{-j \frac{\pi}{2} k \alpha} \mathbf{u}_k[n],
	\label{FRT_eq}
\end{equation}
where $\mathcal{K}\triangleq \{0, 1, \dots, N - 3, N - 2, N-{(N)}_2\} $ and $\mathbf{u}_k$ is the discrete counterpart of the $k^\text{th}$ Hermite-Gaussian function. The peculiar summation range occurs because of the number of zero crossings of the discrete Hermite-Gaussian functions, which are the eigenvectors of the normalized discrete Fourier transform (DFT) matrix. The DFRT is unitary, index additive, and reduces to the identity and normalized DFT when $\alpha=0$ and $\alpha=1$, respectively. Derivation of discrete Hermite-Gaussians and further details regarding the generation of the DFRT matrix can be found in~\cite{candan00dfrt}.

\subsection{Graph Fractional Fourier Transform and Graph Fractional Laplacian}\label{sec:gfrt_gfl}
Let the GFT matrix be $\mathbf{F}_G = \mathbf{P} \mathbf{J}_G \mathbf{P}^{-1}$ be the Jordan decomposition of $\mathbf{F}_G$. There are multiple but equivalent ways of defining the GFRT based on the given form. In our context, the fractional power-based definition is sufficient and given as follows~\cite{wang17gspfrt,alikasifoglu24unified}:
\begin{equation}
	\mathbf{F}_G^\alpha = \mathbf{P} \mathbf{J}_G^\alpha \mathbf{P}^{-1},
	\label{GFRT_eq}
\end{equation}
where $\alpha \in \mathbb{R}$ is the fractional order. In~\cite{alikasifoglu24unified}, we provide a unified theory for multiple but equivalent definitions of GFRT that is applicable to any underlying graph structure and any real fractional order. For additional details, refer to~\cite{alikasifoglu24unified} and its supplementary material. The definition of GFRT is index additive: $\mathbf{F}_G^\alpha \mathbf{F}_G^\beta=\mathbf{F}_G^{\alpha+\beta}$, and also has the properties of reduction to the identity matrix and the GFT matrix when orders are $\alpha=0$ and $\alpha=1$, respectively~\cite{alikasifoglu24unified}.

On the other hand, let $\mathbf{L} = \mathbf{U} \mathbf{\Lambda} \mathbf{U}^H$ be the diagonalization of a graph Laplacian. Then the graph fractional Laplacian of the order $\alpha$ can be defined as~\cite{yan21windowfrftjou}:
\begin{equation}
	\mathbf{L}_{\alpha} = \mathbf{U}^{\alpha} \mathbf{\Lambda}^{\alpha} {(\mathbf{U}^{\alpha})}^{H}.
	\label{fraclap}
\end{equation}
\begin{remark}
	We highlight that the graph fractional Laplacian $\mathbf{L}_{\alpha} = \mathbf{U}^{\alpha} \mathbf{\Lambda}^{\alpha} {(\mathbf{U}^{\alpha})}^{H}$ is a different definition than the fractional power of Laplacian $\mathbf{L}^{\alpha} = \mathbf{U}\mathbf{\Lambda}^{\alpha} \mathbf{U}^{H}$. $\mathbf{L}^\alpha$ is the fractional matrix power, where $\mathbf{L}_\alpha$ is a special definition where the eigenvector matrices also have fractional power.
\end{remark}

\subsection{Joint Time-Vertex Fourier Transform (JFT)}
Let $\mathbf{X} \in \mathbb{C}^{N\times T}$ represent the joint time-vertex signal such that we have graph signals in the columns and time-series signals defined for each vertex of the underlying graph in the rows of $\mathbf{X}$. From the perspective of the columns, $\mathbf{X}$ consists of time-varying graph signals. From the standpoint of the rows, $\mathbf{X}$ contains a time-series signal at each node. The DFT of this signal is given as:
\begin{equation}
	\text{DFT}(\mathbf{X}) = \mathbf{X}\mathbf{F}^T,
\end{equation}
where $\mathbf{F}$ is the normalized DFT matrix with its elements given as $\mathbf{F}_{m,n} = \frac{1}{\sqrt{T}}e^{\frac{-j2\pi m n}{T}}$, $m,n = 0,1, \dots, T-1 $. Similarly, the GFT of $\mathbf{X}$ can be given as:
\begin{equation}
	\text{GFT}(\mathbf{X}; \mathcal{G}) = \mathbf{F}_G\mathbf{X}.
\end{equation}
Finally, the JFT of $\mathbf{X}$ is defined as~\cite{grassi18timevertex,loukas16jft}
\begin{equation}
	\text{JFT}(\mathbf{X}; \mathcal{G}) = \mathbf{F}_G\mathbf{X}\mathbf{F}^T,
\end{equation}
which can capture the spectral information of $\mathbf{X}$ in both time and underlying graph perspectives. To define the transform more compactly, one can denote JFT in matrix form by vectorizing $\mathbf{X}$. Doing so, we have:
\begin{equation}
	\text{JFT}(\mathbf{x}, \mathcal{G}) = \mathbf{F}_J \mathbf{x},
\end{equation}
where $\mathbf{F}_J \triangleq \mathbf{F} \otimes \mathbf{F}_G$ and $\mathbf{x} = vec(\mathbf{X})$.
The inverse JFT is given by:
\begin{equation}
	\text{JFT}^{-1}(\mathbf{X},G) = \mathbf{F}_G^{-1} \mathbf{X} \mathbf{F}^*,
\end{equation}
and $\mathbf{F}_J^{-1} = \mathbf{F}^H \otimes \mathbf{F}_G^{-1}$.
It is shown in~\cite{loukas16jft} that if the underlying $\mathbf{F}_G$ is unitary, so is JFT\@.

\subsection{Joint Time-Vertex Filters}\label{sec:jtv_filter_pre}
A joint time-vertex filter is defined in the joint spectral domain and evaluated at all graph and time frequencies \(\lambda_i\), \(\omega_j\) where \(i=1,\dots, N\), \(j=1,\dots, T\). For a vectorized joint time-vertex signal \(\mathbf{x} \triangleq vec(\mathbf{X})\), the filtered version of the signal, \(\hat{\mathbf{x}}\), is obtained by~\cite{grassi18timevertex}: \(\hat{\mathbf{x}} = \mathbf{F}_J^{-1} \mathbf{H}_J \mathbf{F}_J \mathbf{x}\), where \(\mathbf{F}_J\) and \(\mathbf{F}_J^{-1}\) are joint time-vertex Fourier and inverse Fourier transforms, respectively, and \(\mathbf{H}_J\) is the joint time-vertex filter, whose generic \(NT\times NT\) matrix form is presented, for any function \(h: \mathbb{C}\times\mathbb{R}\to\mathbb{C}\), as follows:
\begin{equation}\label{eq:jtv_filter_generic}
	\mathbf{H}_J = \diag\left \{vec\left(
	\begin{bmatrix}
			h(\lambda_1, \omega_1) & \cdots & h(\lambda_1, \omega_T) \\
			\vdots                 & \ddots & \vdots                 \\
			h(\lambda_N, \omega_1) & \cdots & h(\lambda_N, \omega_T) \\
		\end{bmatrix}
	\right)\right  \}.
\end{equation}
Moreover, these filters are said to be \emph{separable} if \(h(\lambda_i, \omega_j) = h_G(\lambda_i) h_T(\omega_j)\), for suitable functions \(h_G: \mathbb{C}\to\mathbb{C}\) and \(h_T: \mathbb{R}\to\mathbb{C}\), \(\forall i=1,\dots,N\) and \(\forall j=1\dots,T\)~\cite{grassi18timevertex}.

\section{The Joint Time-Vertex Fractional Fourier Transform (JFRT)}\label{sec_frac_jft}
In this section, we define JFRT as a joint generalization of JFT to fractional Fourier and fractional graph Fourier domains. Let $\mathbf{X} \in \mathbb{C}^{N \times T}$ hold a joint time-vertex signal defined on the graph $\mathcal{G}$, where $N$ is the number of vertices of $\mathcal{G}$ and $T$ is the length of the time-series signals defined on each vertex. Then, we define for the order pair $(\alpha, \beta)$ where $\alpha$, $\beta \in \mathbb{R}$, the $(\alpha, \beta)${th} order JFRT as the following:
\begin{equation}\label{defining_jfrt}
	\text{JFT}^{\alpha,\beta}(\mathbf{X}; \mathcal{G}) \triangleq \mathbf{F}_G^{\beta} \mathbf{X}{(\mathbf{F}^{\alpha})}^T,
\end{equation}
where $\mathbf{F}_G^{\beta}$ is the $\beta^\text{th}$ order GFRT defined by~\cref{GFRT_eq} and $\mathbf{F}^{\alpha}$ is the $\alpha^{\text{th}}$ order FRT as defined in~\cref{FRT_eq}. Note that $\mathbf{F}_G$ can be constructed arbitrarily. Also note that~\cref{defining_jfrt}, is a linear transformation from $N \times T$ sized matrices to $N \times T$ sized matrices, that is $\text{JFT}^{\alpha,\beta}(\cdot): \mathbb{C}^{N \times T} \rightarrow \mathbb{C}^{N \times T}$. If we vectorize the given time-vertex signal $\mathbf{X}$ as $\mathbf{x} = vec(\mathbf{X})$, we can find an equivalent transformation between the vectorized forms as $\mathbf{F}_J^{\alpha,\beta}(\cdot): \mathbb{C}^{NT} \rightarrow \mathbb{C}^{NT}$. From the properties of the Kronecker product, such a mapping can be explicitly stated as $\mathbf{F}_J^{\alpha,\beta} \triangleq \mathbf{F}^{\alpha} \otimes \mathbf{F}_G^{\beta}$ in the matrix form.

\subsection{Essential properties of JFRT}
In what follows, we present important propositions and properties of the proposed JFRT operation.
\begin{property}
	For two real-valued pairs $(\alpha_1,\beta_1)$ and $(\alpha_2,\beta_2)$,
	\begin{equation}
		\text{{\normalfont JFT}}^{\alpha_1,\beta_1}(\text{{\normalfont JFT}}^{\alpha_2,\beta_2}(\mathbf{X};\mathcal{G});\mathcal{G}) = \text{{\normalfont JFT}}^{\alpha_1+\alpha_2,\beta_1 + \beta_2}(\mathbf{X},\mathcal{G}).
	\end{equation}
	In other words, JFRT is index additive in fractional orders.
\end{property}
\begin{proof}
	By definition, we have:
	\begin{align*}
		\text{JFT}^{\alpha_1,\beta_1}(\text{JFT}^{\alpha_2,\beta_2}(\mathbf{X};\mathcal{G});\mathcal{G})  = \mathbf{F}_G^{\beta_1} \mathbf{F}_G^{\beta_2} \mathbf{X}{(\mathbf{F}^{\alpha_2})}^T {(\mathbf{F}^{\alpha_1})}^T \\ = \mathbf{F}_G^{\beta_1} \mathbf{F}_G^{\beta_2} \mathbf{X}{(\mathbf{F}^{\alpha_1}\mathbf{F}^{\alpha_2})}^T =  \mathbf{F}_G^{\beta_1+\beta_2} \mathbf{X} {(\mathbf{F}^{\alpha_1+\alpha_2})}^T,
	\end{align*}
	the last equality follows from the index additivity properties of GFRT and FRT\@.
\end{proof}

\begin{property}
	For two real-valued pairs $(\alpha_1,\beta_1)$ and $(\alpha_2,\beta_2)$, JFRT is commutative
	\begin{equation}
		\text{{\normalfont JFT}}^{\alpha_1,\beta_1}(\text{{\normalfont JFT}}^{\alpha_2,\beta_2}(\mathbf{X};\mathcal{G});\mathcal{G}) = \text{{\normalfont JFT}}^{\alpha_2,\beta_2}(\text{{\normalfont JFT}}^{\alpha_1,\beta_1}(\mathbf{X};\mathcal{G});\mathcal{G}).
	\end{equation}
	JFRT is also cross-commutative:
	\begin{align}
		\text{{\normalfont JFT}}^{\alpha_1,\beta_1}(\text{{\normalfont JFT}}^{\alpha_2,\beta_2}(\mathbf{X};\mathcal{G});\mathcal{G}) & = \text{{\normalfont JFT}}^{\alpha_1,\beta_2}(\text{{\normalfont JFT}}^{\alpha_2,\beta_1}(\mathbf{X};\mathcal{G});\mathcal{G}) \nonumber \\
		                                                                                                                             & = \text{{\normalfont JFT}}^{\alpha_2,\beta_1}(\text{{\normalfont JFT}}^{\alpha_1,\beta_2}(\mathbf{X};\mathcal{G});\mathcal{G}).
	\end{align}
\end{property}
\begin{proof}
	The proof follows from the index additivity property:
	\begin{align}
		\text{JFT}^{\alpha_1,\beta_1}(\text{JFT}^{\alpha_2,\beta_2}(\mathbf{X};\mathcal{G});\mathcal{G}) & = \text{JFT}^{\alpha_1 +\alpha_2,\beta_1 +\beta_2}(\mathbf{X};\mathcal{G}) \nonumber                \\
		                                                                                                 & = \text{JFT}^{\alpha_2,\beta_2}(\text{JFT}^{\alpha_1,\beta_1}(\mathbf{X};\mathcal{G});\mathcal{G}).
	\end{align}
	The cross commutativity also follows from index additivity:
	\begin{align}
		\text{JFT}^{\alpha_1,\beta_1}(\text{JFT}^{\alpha_2,\beta_2}(\mathbf{X};\mathcal{G});\mathcal{G}) & = \text{JFT}^{\alpha_1 +\alpha_2,\beta_1 +\beta_2}(\mathbf{X};\mathcal{G}) \nonumber                         \\
		                                                                                                 & = \text{JFT}^{\alpha_1,\beta_2}(\text{JFT}^{\alpha_2,\beta_1}(\mathbf{X};\mathcal{G});\mathcal{G}) \nonumber \\
		                                                                                                 & = \text{JFT}^{\alpha_2,\beta_1}(\text{JFT}^{\alpha_1,\beta_2}(\mathbf{X};\mathcal{G});\mathcal{G}),
	\end{align}
	where the last equality follows from commutativity.
\end{proof}

\begin{property}
	$\text{{\normalfont JFT}}^{0,0}(\mathbf{X};\mathcal{G}) = \mathbf{X}$, which is the reduction to the identity property.
\end{property}
\begin{proof}
	This follows from the definition of JFRT and the reduction to the identity properties of FRT and GFRT\@.
\end{proof}
\begin{property}
	JFRT is separable over graph fractional transform and DFRT, that is, for a given order $(\alpha, \beta)$ we have:
	\begin{equation}
		\text{{\normalfont JFT}}^{\alpha,\beta}(\mathbf{X};\mathcal{G}) = \text{{\normalfont JFT}}^{0,\beta}(\text{{\normalfont JFT}}^{\alpha,0}(\mathbf{X};\mathcal{G})) = \text{{\normalfont JFT}}^{0,\beta}(\text{{\normalfont JFT}}^{\alpha,0} (\mathbf{X};\mathcal{G})).
		\label{separability}
	\end{equation}
\end{property}
\begin{proof}
	First, equality follows from the index additivity property, while the second follows from the commutativity property. Notice that from the reduction to the identity properties of GFRT and DFRT, and using the definition of JFRT, we have the equation~\cref{separability} as:
	\begin{equation}
		\mathbf{F}_{G}^{\beta} \mathbf{X} \mathbf{F}^{\alpha} = \mathbf{F}_{G}^{\beta} (\mathbf{I}_N \mathbf{X} \mathbf{F}^{\alpha})\mathbf{I}_T = \mathbf{I}_{N} (\mathbf{F}_{G}^{\beta} \mathbf{X} \mathbf{I}_{T})\mathbf{F}^{\alpha},
	\end{equation}
	which clearly shows the separable nature of JFRT\@.
\end{proof}
\begin{property}
	For an order pair $(\alpha,\beta)$, JFRT is reversible: \[\text{{\normalfont JFT}}^{-\alpha,-\beta}(\text{{\normalfont JFT}}^{\alpha,\beta}(\mathbf{X};\mathcal{G});\mathcal{G}) = \mathbf{X}.\]
\end{property}
\begin{proof}
	This follows from the index additivity and the reduction to the identity properties of JFRT\@.
\end{proof}

\begin{property}
	Reduction to the ordinary transformation as given by $\text{{\normalfont JFT}}^{1,1}(\mathbf{X};\mathcal{G}) = \text{{\normalfont  JFT}}(\mathbf{X};\mathcal{G})$.
\end{property}
\begin{proof}
	This also follows from the definition of JFRT and the reduction to the ordinary transform properties of FRT and GFRT\@. Thus, JFT becomes a special case of JFRT when order $(1,1)$.
\end{proof}

\begin{proposition}
	If $\mathbf{F}_G$ is a unitary transformation, then so is $\text{{\normalfont JFT}}^{\alpha,\beta}(\mathbf{X;\mathcal{G}})$ or equivalently $\mathbf{F}_J^{\alpha,\beta}$ is a unitary matrix.
\end{proposition}

\begin{proof}
	From the properties of the Kronecker product, we know that if $\mathbf{F}^{\alpha}$ and $\mathbf{F}_G^{\beta}$ are unitary matrices, then so is $\mathbf{F}_J^{\alpha,\beta}$. We know that $\mathbf{F}^{\alpha}$ is unitary for any $\alpha$ by the properties of FRT~\cite{ozaktas01book}. Hence, we only need to show that if $\mathbf{F}_G$ is unitary, then $\mathbf{F}_G^{\beta}$ is also unitary for any $\beta$. If $\mathbf{F}_G$ is unitary, it can be diagonalized as $\mathbf{F}_G = \mathbf{V}_G \mathbf{\Lambda}_G \mathbf{V}_G^H$ with unitary $\mathbf{V}_G$ and diagonal $\mathbf{\Lambda}_G$ with eigenvalues lying on the unit circle. Then, $\mathbf{F}_G^{\beta} = \mathbf{V}_G \mathbf{\Lambda}_G^{\beta} \mathbf{V}_G^H$ will also be unitary since the diagonal values of $\mathbf{\Lambda}_G^{\beta}$ will be on the unit circle as well.
\end{proof}
Hence, for unitary $\mathbf{F}_G$, JFRT is also a unitary transform satisfying the Parseval's relation. On the other hand, it is known that for directed circular graphs, DFT diagonalizes the adjacency matrix, and for an undirected ring graph, DFT diagonalizes the Laplacian matrix. Therefore, for these particular graphs, the GFT can be taken as the DFT matrix, or we can perceive time-series data as a graph in the shape of a directed circular graph or undirected ring graph~\cite{sandryhaila13discretegsp}. Then, we have the following proposition:
\begin{proposition}
	If the underlying graph is a directed circular graph or ring graph, then $\text{{\normalfont JFT}}^{\alpha, \beta}(\mathbf{X; \mathcal{G}})$ reduces to two-dimensional DFRT with orders $\alpha$ and $\beta$.
\end{proposition}

\begin{proof}
	We have $\text{JFT}^{\alpha, \beta}(\mathbf{X}; \mathcal{G}) = \mathbf{F}_G^{\beta} \mathbf{X} {(\mathbf{F}^{\alpha})}^T$.
	Then, one can reduce $\mathbf{F}_G$ to $\mathbf{F}$ due to the underlying graph as follows: From~\cref{FRT_eq}, it can be seen that for even $N$,
	\begin{equation*}
		\mathbf{F}_G = \mathbf{V}_F\, \diag \{e^{-j\frac{\pi}{2}0}, e^{-j\frac{\pi}{2}1}, \dots, e^{-j\frac{\pi}{2}(N-1)} \} \, \mathbf{V}_F^H,
	\end{equation*}
	is a unitary diagonalization of \(\mathbf{F}\), for \(\mathbf{V}_F = [\mathbf{u}_0\, \mathbf{u}_1\, \cdots \, \mathbf{u}_{N-1}]\), where \(\mathbf{u}_k\) is $k^\text{th}$ discrete Hermite-Gaussians. Then, we have
	\begin{equation}
		\mathbf{F}_G^{\beta} =  \mathbf{V}_F\, \diag \{e^{-j\frac{\pi}{2}0 \beta}, \dots, e^{-j\frac{\pi}{2}(N-1)\beta}\} \, \mathbf{V}_F^H = \mathbf{F}^{\beta}.
	\end{equation}
	A similar diagonalization can also be made for odd $N$. Hence, we obtain $\text{JFT}^{\alpha, \beta}(\mathbf{X}; \mathcal{G}) = \mathbf{F}^{\beta} \mathbf{X} {(\mathbf{F}^{\alpha})}^T$ which is equivalent to the two-dimensional DFRT of $\mathbf{X}$.
\end{proof}
\begin{corollary}
	If the underlying graphs are either directed circular graph or ring graph, then $\text{{\normalfont JFT}}^{1,1}(\mathbf{X},\mathcal{G})$ can be represented as two-dimensional DFT\@.
\end{corollary}
\begin{proof}
	This follows from the previous proposition and reduction to the ordinary transform property of FRT\@.
\end{proof}

As a two-dimensional transformation, JFRT has the properties of index additivity, reduction to the identity, and reduction to the ordinary transformations, just as the two-dimensional DFRT\@. It also has the unitarity property provided that $\mathbf{F}_G$ is unitary. This condition is satisfied for undirected graphs with common GFT definitions~\cite{sandryhaila13discretegsp,shuman13emerging}. Also, there are unitary transformations for the directed graphs introduced by recent works~\cite{shafipour18digraphconf}. Therefore, JFRT is a good candidate that generalizes the discrete manifestation of the classical signal processing's two-dimensional FRT (2D-DFRT) to the GSP domain.

\subsection{Fractional Joint Time-Vertex Filters}\label{sec:jtv_frac_filter_theory}
Joint time-vertex filters are mentioned in Section~\ref{sec:jtv_filter_pre}, as \(NT\times NT\) matrices, \(\mathbf{H}_J\). In this section, we extend this notion to \emph{the fractional joint time-vertex filters}, referred to as \(\mathbf{H}_J^{(\alpha,\beta)}\), where we use fractional order pairs \((\alpha,\beta)\) as superscripts to refer graph fraction of \(\beta \) and time fraction of \(\alpha \). A fractional joint time-vertex filter \(\mathbf{H}_J^{(\alpha,\beta)}\) has the generic form presented as:
\begin{equation}\label{eq:jtv_frac_filter_generic}
	\scalemath{0.95}{
	\mathbf{H}_J^{(\alpha,\beta)} = \diag\left \{vec\left(
	\begin{bmatrix}
			h(\lambda_1^\beta, \omega_1^\alpha) & \cdots & h(\lambda_1^\beta, \omega_T^\alpha) \\
			\vdots                              & \ddots & \vdots                              \\
			h(\lambda_N^\beta, \omega_1^\alpha) & \cdots & h(\lambda_N^\beta, \omega_T^\alpha) \\
		\end{bmatrix}
	\right)\right \},
	}
\end{equation}
where the special case for \((\alpha, \beta)=(1,1)\) reduces to~\cref{eq:jtv_filter_generic}. With this definition, the joint time-vertex filtering can be conducted for a joint time-vertex signal \(\mathbf{x} \triangleq vec(\mathbf{X})\) by:
\begin{equation}
	\hat{\mathbf{x}} = \mathbf{F}_J^{(-\alpha,-\beta)} \mathbf{H}_J^{(\alpha,\beta)} \mathbf{F}_J^{(\alpha,\beta)} \mathbf{x}.
\end{equation}

\subsubsection{Separable Fractional Joint Time-Vertex Filters}\label{sec:seperable_frac_jtv_filter_theory}
Separable joint time-vertex filters presented in Section~\ref{sec:jtv_filter_pre} can also be generalized to fractional joint time-vertex filters as the filters that satisfy \(h(\lambda_i^\beta,\omega_j^\alpha) = h_G(\lambda_i^\beta) h_T(\omega_j^\alpha)$, $\forall i=1,\dots, N,\forall j=1,\dots, T\) and \(\forall \alpha,\beta\in\mathbb{R}\). With this definition, a separable joint time-vertex filter \(\mathbf{H}_J^{(\alpha,\beta)}\) can be represented as a Kronecker product of a graph filter \(\mathbf{H}_G^\beta \) and a time filter \(\mathbf{H}_T^\alpha \) as \(\mathbf{H}_J^{(\alpha,\beta)} = \mathbf{H}_T^\alpha \otimes \mathbf{H}_G^\beta \). Next, we show that transforming and filtering operations can be conducted on fractional graph spectral and fractional frequency domains, separately, for a \emph{separable} fractional joint time-vertex filter \(\mathbf{H}_J^{(\alpha,\beta)}\):
\begin{proposition}
	For a separable fractional joint time-vertex filter \(\mathbf{H}_J^{(\alpha,\beta)}\) in the form of \(\mathbf{H}_J^{(\alpha,\beta)} = \mathbf{H}_T^\alpha \otimes \mathbf{H}_G^\beta \), and a joint time-vertex signal \(\mathbf{X}\in\mathbb{C}^{N\times T}\) whose fractional joint time-vertex filtered version \(\hat{\mathbf{X}}\) can be obtained through:
	\begin{equation}\label{eq:prop:jtv_frac_filter_vec}
		vec(\hat{\mathbf{X}}) = \mathbf{F}_J^{(-\alpha,-\beta)} \mathbf{H}_J^{(\alpha,\beta)} \mathbf{F}_J^{(\alpha,\beta)} vec(\mathbf{X}),
	\end{equation}
	there is a graph domain transformation \(\mathbf{T}_G\) and a time-domain transformation \(\mathbf{T}_T\) such that \(\hat{\mathbf{X}} = \mathbf{T}_G \mathbf{X} \mathbf{T}_T\).
\end{proposition}
\begin{proof}
	By using the relation of the \(vec(\cdot)\) operator and the Kronecker product (\(\otimes \)) continuously, we can write~\cref{eq:prop:jtv_frac_filter_vec} as:
	\begin{align}
		vec(\hat{\mathbf{X}}) & = \mathbf{F}_J^{(-\alpha,-\beta)} \mathbf{H}_J^{(\alpha,\beta)} \mathbf{F}_J^{(\alpha,\beta)} vec(\mathbf{X})                                                                                                                                                  \\
		                      & = \mathbf{F}_J^{(-\alpha,-\beta)} \mathbf{H}_J^{(\alpha,\beta)}  vec(\mathbf{F}_G^{\beta}\mathbf{X}{\left(\mathbf{F}^{\alpha}\right)}^T)                                                                                                                       \\
		                      & = \mathbf{F}_J^{(-\alpha,-\beta)} vec(\mathbf{H}_G^{\beta}\mathbf{F}_G^{\beta}\mathbf{X}{\left(\mathbf{F}^{\alpha}\right)}^T{\left(\mathbf{H}_T^{\alpha}\right)}^T)                                                                                            \\
		                      & = vec\Big(\underbrace{\mathbf{F}_G^{-\beta}\mathbf{H}_G^{\beta}\mathbf{F}_G^{\beta}}_{\mathbf{T}_G}\mathbf{X}\underbrace{{\left(\mathbf{F}^{\alpha}\right)}^T{\left(\mathbf{H}_T^{\alpha}\right)}^T{\left(\mathbf{F}^{-\alpha}\right)}^T}_{\mathbf{T}_T}\Big).
	\end{align}
	Therefore, \(\hat{\mathbf{X}} = \mathbf{T}_G\mathbf{X}\mathbf{T}_T\).
\end{proof}

\subsection{Computational Cost Analysis}
Since both JFT and JFRT are separable, their computational complexity analysis can be done separately for time-domain and graph-domain transformations. In JFT, we take the DFT of $N$ rows in $\mathcal{O} (N T \log T)$ complexity using the fast Fourier transform (FFT). Similarly, for a $T$ length vector, its DFRT can also be efficiently computed using the established fast algorithm in~\cite{ozaktas96fastFRT}. Hence, taking DFRT of $N$ rows has the complexity of $\mathcal{O}(N T \log T)$ as well. For graph transformations, the computational cost of taking a GFT of a vector of length $N$ is $\mathcal{O} (N^2)$ since it is a matrix-vector multiplication. Although there are works that reduce the computational cost in obtaining exact or approximate transforms~\cite{magoarou16approximatefastgft,domingos20gft,lu19fastgft} for certain classes of graphs, there are no generalized fast GFT computation algorithms like the FFT and the fast DFRT\@. Since GFRT is also in a matrix form, transforming $T$ columns is of complexity $\mathcal{O} (T N^2)$ for both GFT and GFRT\@. This gives a total complexity of $\mathcal{O} (NT(\log T + N))$ for calculating either JFT or JFRT\@. Thus, JFRT brings no extra computational burden in terms of transformation complexity. On the other hand, to obtain a transformation matrix for the Laplacian approach, one needs to calculate the diagonalization of the graph Laplacian matrix in $\mathcal{O} (N^3)$ time, and for the algebraic signal processing approach, one may need to calculate the Jordan decomposition of the adjacency matrix, which is even more computationally expensive than the diagonalization cost. However, these calculations need to be performed only once for a particular underlying graph. Hence, it should not be confused with transformation complexity for successive signal operations.

For JFRT filtering, let $\mathbf{X} \in \mathbb{C}^{N \times T}$ be a joint time-vertex signal and let the graph Laplacian of the underlying graph be $\mathbf{L}$, and consider filter $h(\lambda, \omega)$ with $\lambda\in \{ \lambda_1, \dots \lambda_N \}$, $\omega \in \{ \omega_1, \dots, \omega_T \}$. For the ordinary JFT, the FFT of every row is first computed with the computational complexity $\mathcal{O}(NT\log T )$. Then for each column of $\mathbf{X}$, a $K${th} order Chebyshev approximation of $h(\lambda, \omega_k)$, where $\omega_k = 2\pi k/T$, $k = 0,1, \dots, T-1$, is computed with the computational complexity $\mathcal{O} (KT|\mathcal{E}|)$, where $\mathcal{E}$ is the set of edges for the underlying graph~\cite{grassi18timevertex}. Then, the inverse FFT of each row is calculated with complexity $\mathcal{O} (NT \log T)$. For JFRT filtering, one again looks at $\mathcal{O}(NT\log T )$ complexity for taking DFRTs of every row since DFRTs can also be efficiently computed in $\mathcal{O}(T\log T)$ time~\cite{ozaktas96fastFRT}. Then, we filter each column of $\mathbf{X}$ with $h_{frac}(\lambda,\omega)$. Since we have already computed the diagonalization of $\mathbf{L}$ to obtain GFRT, we can filter each column of $\mathbf{X}$ using the already obtained GFRT\@. Hence, effectively, we reach the total computational complexity of $\mathcal{O}(N^2 T)$ for the computation of GFRT filtering of each column. Lastly, the inverse DFRT of each column is calculated with $\mathcal{O}(NT\log T )$ complexity.

\section{Tikhonov Regularized Denoising with JFRT}\label{tiknovossec}
Here, we develop the underlying theory for Tikhonov regularization-based denoising by the proposed JFRT\@. We first present some properties of fractional Laplacians and define joint fractional Laplacian. Then, we introduce a fractional variation measure and show the connection of Tikhonov regularization with joint fractional Laplacians. These are used to derive regularization-based denoising by JFRT and to find the optimal filter in JFRT domains.

\subsection{Graph, Time, and Joint Fractional Laplacians} Let us first remember the edge derivative and graph gradient definitions. In the graph Laplacian approach, for a graph signal $\mathbf{x} \in \mathbb{R}^N$, the edge derivative with respect to the edge $e = (m,n)$ at vertex $m$ can be given as~\cite{shuman13emerging,loukas16jft}:
\begin{equation}
	\left. \frac{\partial \mathbf{x}}{\partial e} \right|_{m} = \sqrt{\mathbf{A}_{m,n}} [\mathbf{x}_m - \mathbf{x}_n], \end{equation} where $\mathbf{A}$ is the adjacency matrix of the given graph. Then the graph gradient of $\mathbf{x}$ at vertex $m$ can also be given as~\cite{shuman13emerging,loukas16jft}:
\begin{equation}
	\left. \nabla_{G} \mathbf{x}_m \right|_{m} = \Big {\{ \left.
	\frac{\partial \mathbf{x}}{\partial e} \right|_{m} \Big \}}_{e \in \mathcal{E}}.
\end{equation}
Without the loss of generality, once an ordering of edges is set, the graph gradient can be represented as a matrix $\bm{\nabla}_{G} \in \mathbb{R}^{|\mathcal{E}| \times N}$ with elements:
\begin{equation}\label{circular_def}
	{(\bm{\nabla}_{G})}_{k,m} =
	\begin{cases}
		\phantom{-}\sqrt{\mathbf{A}_{m,n}}, & \text{ if } e_k = (m,n), \\
		-\sqrt{\mathbf{A}_{m,n}},           & \text{ if } e_k = (n,m), \\
		0,                                  & \text{ otherwise},
	\end{cases}
\end{equation}
where $e_k$ is the $k^\text{th}$ edge where $k = 1,\ldots,|\mathcal{E}|$. Then the graph divergence of the graph gradient gives the graph Laplacian as $\mathbf{L}_{G} = {(\bm{\nabla}_{G})}^T (\bm{\nabla}_{G})$.
Similarly, a $T$-dimensional time-series signal can be considered a circular ring graph~\cite{sandryhaila13discretegsp,shuman13emerging}.
For circular ring graphs, we have:
\begin{equation}
	\mathbf{(A_{circ})}_{m,n} = \begin{cases}
		1 \text{ if } {(m-n)}_T= 1, \\
		0, \text{otherwise}.
	\end{cases}
\end{equation}
Let $e_1 = (1,2)$, $e_2 = (2,3)$, \dots, $e_{T-1} = (T-1,T)$, $e_{T} = (T,1)$, then let the graph gradient of circular ring graph be $\bm{\nabla}_{T}$. Notice that $\bm{\nabla}_{T}$ is the first order difference operator such that ${(\bm{\nabla}_{T} \mathbf{x})}_m= \mathbf{x}_m - \mathbf{x}_{{(m-1)}_N} $. We then finally have the time Laplacian as $\mathbf{L}_{T} = {(\bm{\nabla}_{T})}^T \bm{\nabla}_{T}$~\cite{shuman13emerging,grassi18timevertex}. For a fractional order $\alpha$, as we have given in Section~\ref{sec:gfrt_gfl}, we have the graph fractional Laplacian as ${(\mathbf{L_G})}_{\alpha} = \mathbf{U}_G^{\alpha} \mathbf{\Lambda}^{\alpha} {(\mathbf{U}_G^{\alpha})}^H$. It is known that the graph Laplacian is a positive semi-definite Hermitian matrix~\cite{shuman13emerging}. Next, we show that the graph fractional Laplacian is also a positive semi-definite Hermitian matrix:
\begin{proposition}\label{proposition_possemi}
	For any fractional order $\alpha$, ${(\mathbf{L_G})}_{\alpha}$ is a positive semi-definite Hermitian matrix.
\end{proposition}
\begin{proof}
	We have ${(\mathbf{L_G})}_{\alpha} = \mathbf{U}_G^{\alpha} \mathbf{\Lambda}^{\alpha} {(\mathbf{U}_G^{\alpha})}^H$. Define $\mathbf{M}_G = \mathbf{\Lambda}^{\alpha/2} {(\mathbf{U}_G^{\alpha})}^H$.
	Notice that since ${(\mathbf{L_G})}_{1} = \mathbf{L}_G$ is Hermitian, all of the elements of $\mathbf{\Lambda}$ are real. Then ${(\mathbf{L_G})}_{\alpha} = \mathbf{M}_G^H \mathbf{M}_G$ is Hermitian (by construction) and positive semi-definite since for any $\mathbf{x} \in \mathbb{C}^{N}$ we have:
	\begin{equation*}
		\mathbf{x}^H {(\mathbf{L_G})}_{\alpha} \mathbf{x} = \mathbf{x}^H \mathbf{M}_G^H \mathbf{M}_G\mathbf{x} =  {(\mathbf{M}_G\mathbf{x})}^H \mathbf{M}_G\mathbf{x} \geq 0.
	\end{equation*}
\end{proof}
Now, recall that the diagonalizations of the time and graph fractional Laplacians are given as ${(\mathbf{L}_T)}_{\alpha} = \mathbf{U}_T^{\alpha} \mathbf{\Lambda}_T^{\alpha} {(\mathbf{U}_T^{\alpha})}^H$ and ${(\mathbf{L}_G)}_{\beta} = \mathbf{U}_G^{\beta} \mathbf{\Lambda}_G^{\beta} {(\mathbf{U}_G^{\beta})}^H$ for fractional orders $\alpha$ and $\beta$, respectively. We can now define the joint fractional Laplacian $\mathbf{L}_J^{(\alpha, \beta)}$ as the Kronecker sum of the time and graph fractional Laplacians parallel to the construction of the ordinary joint Laplacian~\cite{grassi18timevertex}, note that we use \((\alpha,\beta)\) as superscripts:
\begin{align}
	\mathbf{L}_J^{(\alpha, \beta)} & \triangleq {(\mathbf{L}_T)}_{\alpha} \oplus {(\mathbf{L}_G)}_{\beta} = {(\mathbf{L}_T)}_{\alpha} \otimes \mathbf{I}_G + \mathbf{I}_T \otimes {(\mathbf{L}_G)}_{\beta} \nonumber           \\ &=(\mathbf{U}_T^{\alpha} \mathbf{\Lambda}_T^{\alpha}
	{(\mathbf{U}_T^{\alpha})}^H) \otimes \mathbf{I}_G + \mathbf{I}_T \otimes (\mathbf{U}_G^{\beta} \mathbf{\Lambda}_G^{\beta}
	{(\mathbf{U}_G^{\beta})}^H) \nonumber                                                                                                                                                                                      \\
	                               & = (\mathbf{U}_T^{\alpha} \otimes \mathbf{U}_G^{\beta}) (\mathbf{\Lambda}_T^{\alpha} \oplus \mathbf{\Lambda}_G^{\beta}) {(\mathbf{U}_T^{\alpha} \otimes \mathbf{U}_G^{\beta})}^H \nonumber \\ &= \mathbf{U}_J^{(\alpha,\beta)} \mathbf{\Lambda}_J^{(\alpha,\beta)} {(\mathbf{U}_J^{(\alpha,\beta)})}^H.
	\label{kronecker_sum_laplacian}
\end{align}
Noticing ${(\mathbf{U}_J^{(\alpha,\beta)})}^H = {(\mathbf{U}_T^{\alpha} \otimes \mathbf{U}_G^{\beta})}^H = \mathbf{F}^{\alpha} \otimes \mathbf{F}_G^{\beta} = \mathbf{F}_J^{(\alpha, \beta)}$, the eigenvector matrices obtained from the diagonalization of the joint fractional Laplacian form a basis for the JFRT\@. Conversely, JFRT diagonalizes the joint fractional Laplacian.

\begin{proposition}\label{proposition_possemi_2d}
	The joint fractional Laplacian is a positive semi-definite Hermitian matrix for any fractional pair $(\alpha, \beta)$; the joint fractional Laplacian is a positive semi-definite Hermitian matrix.
\end{proposition}
\begin{proof}
	For the Hermitian, by the definition given in~\cref{kronecker_sum_laplacian}, we need to only show that $\mathbf{\Lambda}_J^{(\alpha,\beta)}$ is Hermitian. From Proposition~\ref{proposition_possemi}, $\mathbf{\Lambda}_T^{\alpha}$ and $\mathbf{\Lambda}_G^{\beta}$ are non-negative real-valued diagonal matrices. In~\cref{kronecker_sum_laplacian}, $\mathbf{\Lambda}_J^{(\alpha,\beta)} = (\mathbf{\Lambda}_T^{\alpha} \oplus \mathbf{\Lambda}_G^{\beta})$ is also real-valued diagonal matrix due to the properties of the Kronecker sum, hence it is Hermitian. From the diagonalization of $\mathbf{L}_J^{(\alpha, \beta)}$ in~\cref{kronecker_sum_laplacian}, Hermitian property follows. Let $\mathbf{M}_J \triangleq {(\mathbf{\Lambda}_J^{(\alpha,\beta)})}^{0.5} {(\mathbf{U}_J^{(\alpha,\beta)})}^H $, for positive semi-definiteness. Then we have $\mathbf{L}_J^{(\alpha, \beta)} = \mathbf{M}_J^H \mathbf{M}_J$. Thus, for any $\mathbf{x} \in \mathbb{C}^{NT}$ we have:
	\begin{equation*}
		\mathbf{x}^H \mathbf{L}_J^{(\alpha, \beta)} \mathbf{x} = \mathbf{x}^H \mathbf{M}_J^H \mathbf{M}_J \mathbf{x} = {(\mathbf{M}_J \mathbf{x})}^H (\mathbf{M}_J \mathbf{x}) \geq 0.
	\end{equation*}
\end{proof}
\noindent We also provide some properties of JFT that will be used later. First, we have the following relation for JFT~\cite{grassi18timevertex}:
\begin{equation}
	\bm{\nabla}_J \mathbf{x} = vec \Big (
	\begin{bmatrix}
			\mathbf{X} {(\bm{\nabla}_T)}^T \\
			(\bm{\nabla}_G) \mathbf{X}
		\end{bmatrix}
	\Big )\text{ for }
	\bm{\nabla}_J=
	\begin{bmatrix}
		(\bm{\nabla}_T) \otimes \mathbf{I}_G \\
		\mathbf{I}_T \otimes (\bm{\nabla}_G)
	\end{bmatrix},
\end{equation}
where $\bm{\nabla}_J$ is the joint gradient. Second, using~\cref{kronecker_sum_laplacian} with order $(1,1)$, it can be shown that $\bm{\nabla}_J^T \bm{\nabla}_J = ({(\bm{\nabla}_T)}^T (\bm{\nabla}_T)) \otimes \mathbf{I}_G +\mathbf{I}_T \otimes {(\bm{\nabla}_G)}^T ( \bm{\nabla}_G) =  (\mathbf{L}_T) \otimes \mathbf{I}_G + \mathbf{I}_T \otimes (\mathbf{L}_G) = \mathbf{L}_J^{(1,1)} = \mathbf{L}_J $.
We also define the joint fractional variation as the following:
\begin{align}
	\mathbf{x}^H \mathbf{L}_J^{(\alpha, \beta)} \mathbf{x} = \mathbf{x}^H ({(\mathbf{L}_T)}_{\alpha} \oplus {(\mathbf{L}_G)}_{\beta}) \mathbf{x}, \label{fractional_var}
\end{align}
which measures the variation of the input joint time-vertex signal $\mathbf{x}$ with respect to the JFRT modes that ${(\mathbf{L}_T)}_{\alpha}$ and ${(\mathbf{L}_G)}_{\beta}$ constitutes.
Notice that for order (1,1), this reduces to the Laplacian quadratic form with ordinary joint Laplacian $\mathbf{L}_J^{(1,1)} = \mathbf{L}_J$~\cite{shuman13emerging}.  Since $\mathbf{L}_J^{(\alpha, \beta)} $ is positive semi-definite from Proposition~\ref{proposition_possemi_2d},~\cref{fractional_var} gives non-negative real values.

\subsection{Tikhonov Regularization with Joint Fractional Variation}\label{denoise_theory}
We have the following joint time-vertex signal model in the Tikhonov regularization based denoising: \(\mathbf{Y} = \mathbf{X} + \mathbf{N}\),  where $\mathbf{y} = vec(\mathbf{Y}) $ is the received signal, $\mathbf{x} = vec(\mathbf{X})$ is the input signal, and $\mathbf{n} = vec(\mathbf{N})$ is the i.i.d. Gaussian noise distributed across two domains. For JFT, one way to denoise the signal $\mathbf{y}$ is to minimize $l_2$ regularized objective function~\cite{grassi18timevertex,shuman13emerging}:
\begin{equation}
	\label{Tikhonov_opt_regular}
	\mathbf{\hat{x}}= \argmin_{\mathbf{x} = vec(\mathbf{X})}  \Vert  \mathbf{y} - \mathbf{x}  \Vert_2^2 + \tau_g \Vert
	{(\bm{\nabla})}_G \mathbf{X} \Vert_F^2 + \tau_t \Vert
	\mathbf{X}{(\bm{\nabla})}_T \Vert_F^2,
\end{equation}
where $\tau_g$ and $\tau_t$ are non-negative real regularization parameters. One can show that the regularization component of~\cref{Tikhonov_opt_regular} can be written as a Laplacian quadratic form:
\begin{align}
	\tau_g \Vert  {(\bm{\nabla})}_G \mathbf{X} \Vert_F^2 + \tau_t \Vert  \mathbf{X}{(\bm{\nabla})}_T \Vert_F^2\,     & = \Vert  \sqrt{ \tau_g} {(\bm{\nabla})}_G \mathbf{X} \Vert_F^2 + \Vert  \mathbf{X} \sqrt{\tau_t}{(\bm{\nabla})}_T \Vert_F^2 \nonumber \\
	=\Vert  \bm{\nabla}_J^{\tau_g, \tau_t} \mathbf{x}  \Vert_2^2 = \mathbf{x}^H {(\bm{\nabla}_J^{\tau_g, \tau_t})}^T & (\bm{\nabla}_J^{\tau_g, \tau_t}) \mathbf{x} = \mathbf{x}^H \mathbf{L}_J^{\tau_g, \tau_t}  \mathbf{x},
	\label{equivalent_reg}
\end{align}
where $\bm{\nabla}_J^{\tau_g, \tau_t}$ denotes the regularized joint gradient as:
\begin{equation}
	\bm{\nabla}_J^{\tau_g, \tau_t} \triangleq
	\begin{bmatrix}
		(\sqrt{\tau_t} \bm{\nabla}_T) \otimes \mathbf{I}_G \\
		\mathbf{I}_T \otimes (\sqrt{\tau_g}\bm{\nabla}_G)
	\end{bmatrix},
\end{equation}
and the regularized joint Laplacian is given as:
\begin{equation}
	\mathbf{L}_J^{\tau_g, \tau_t} \triangleq {(\bm{\nabla}_J^{\tau_g, \tau_t})}^T (\bm{\nabla}_J^{\tau_g, \tau_t}).
\end{equation}
Second, since
\begin{align}
	\mathbf{L}_J^{\tau_g, \tau_t} & = {(\bm{\nabla}_J^{\tau_g, \tau_t})}^T (\bm{\nabla}_J^{\tau_g, \tau_t}) = {((\sqrt{\tau_t}\bm{\nabla}_T) \otimes \mathbf{I}_G)}^T (( \sqrt{\tau_t}\bm{\nabla}_T) \otimes \mathbf{I}_G) \nonumber                   \\
	                              & \phantom{= {(\bm{\nabla}_J^{\tau_g, \tau_t})}^T (\bm{\nabla}_J^{\tau_g, \tau_t}) \: \, } + {(\mathbf{I}_T \otimes (\sqrt{\tau_g} \bm{\nabla}_G))}^T  (\mathbf{I}_T \otimes (\sqrt{\tau_g}\bm{\nabla}_G)) \nonumber \\
	                              & = ({(\sqrt{\tau_t}\bm{\nabla}_T)}^T (\sqrt{\tau_t} \bm{\nabla}_T)) \otimes \mathbf{I}_G +\mathbf{I}_T \otimes ({(\sqrt{\tau_g} \bm{\nabla}_G)}^T (\sqrt{\tau_g} \bm{\nabla}_G)) \nonumber                          \\
	                              & = (\tau_t \mathbf{L}_T) \otimes \mathbf{I}_G + \mathbf{I}_T \otimes (\tau_g \mathbf{L}_G) = \tau_t (\mathbf{L}_T) \oplus \tau_g(\mathbf{L}_G),
	\label{modified_definition}
\end{align}
the regularized joint Laplacian can be written as the Kronecker sum of $\tau_t$-scaled time Laplacian and $\tau_g$-scaled graph Laplacian. By applying the definition of the joint fractional Laplacian given in~\cref{kronecker_sum_laplacian} to the result obtained in~\cref{modified_definition}, we can define a regularized joint fractional Laplacian ${(\mathbf{L}_J^{\tau_g,\tau_t})}^{(\alpha, \beta)}$ by replacing ordinary time and graph Laplacians with fractional ones:
\begin{equation}
	{(\mathbf{L}_J^{\tau_g,\tau_t})}^{(\alpha, \beta)} \triangleq  \tau_t {(\mathbf{L}_T)}_{\alpha} \oplus \tau_g {(\mathbf{L}_G)}_{\beta}.
\end{equation}
Now, we are in a position to use the regularized joint fractional Laplacian ${(\mathbf{L}_J^{\tau_g,\tau_t})}^{(\alpha, \beta)}$ in parallel to the result obtained in~\cref{equivalent_reg} to do Tikhonov regularization across both graph fractional and time fractional domains by plugging it in~\cref{fractional_var} as a regularization parameter. Thus, we can write the  regularization-based denoising objective function for JFRT as:
\begin{align}
	\mathbf{\hat{x}}= \argmin_{\mathbf{x} = vec(\mathbf{X})}  \Vert \left. \mathbf{y} - \mathbf{x} \right. \Vert_2^2 +  \mathbf{x}^H {(\mathbf{L}_J^{\tau_g,\tau_t})}^{(\alpha, \beta)} \mathbf{x}.
	\label{Tikhonov_opt_modified}
\end{align}
The minimum value of~\cref{Tikhonov_opt_modified} can be found as a filter in the JFRT domain, as shown in the following proposition:
\begin{proposition}\label{proposition_opt_mul_filter}
	$\mathbf{\hat{x}} = \mathbf{F}_J^{(-\alpha,-\beta)} \mathbf{H}_J^{(\alpha,\beta)} \mathbf{F}_J^{(\alpha,\beta)} \mathbf{y}$ is the minimizer in~\cref{Tikhonov_opt_modified} where $\mathbf{H}_J^{(\alpha,\beta)}$ is the optimal multiplicative filter with diagonal elements given by:
	\begin{equation}
		\diag(\mathbf{H}_J^{(\alpha,\beta)}) = vec(\mathbf{H}^{(\alpha,\beta)}).
	\end{equation}
	$\mathbf{H}^{(\alpha,\beta)}$ is the matrix manifestation of \begin{equation}
		h^{(\alpha,\beta)}_{m,n} = \frac{1}{1+\tau_g \lambda_m^{\beta} + \tau_t \omega_n^{\alpha}}, \end{equation} where $\lambda_m$ and $\omega_n$ are the eigenvalues of graph and time Laplacians with $m = 1,2,\dots, N$ and $n = 1,2,\dots,T$, respectively.
\end{proposition}
\begin{proof}
	Let $\mathbf{x} = \mathbf{x}_r + j \mathbf{x}_j$ where $\mathbf{x}_r$ and $\mathbf{x}_j$ are real valued vectors, and $j$ is the imaginary unit. Then, since~\cref{Tikhonov_opt_modified} is a convex function of $\mathbf{x}$, we can take the derivatives with respect to $\mathbf{x}_r$ and $\mathbf{x}_j$ and set them to zero. Let $f(\mathbf{x})$ be the argument of~\cref{Tikhonov_opt_modified}, i.e. $f(\mathbf{x}) = {(\mathbf{y} - \mathbf{x})}^H (\mathbf{y} - \mathbf{x}) + \mathbf{x}^H {(\mathbf{L}_J^{\tau_g,\tau_t})}^{(\alpha, \beta)} \mathbf{x}$. By knowing that ${(\mathbf{L}_J^{\tau_g,\tau_t})}^{(\alpha, \beta)}$ is an Hermitian matrix by Proposition~\ref{proposition_possemi_2d}, setting $\frac{\partial f(\mathbf{x})}{\partial \mathbf{x}_r} = 0$ gives:
	\begin{equation}
		\Re \{\mathbf{y}\} =  \mathbf{x}_r + \Re \{ {(\mathbf{L}_J^{\tau_g,\tau_t})}^{(\alpha, \beta)}\} \mathbf{x}_r - \Im \{ {(\mathbf{L}_J^{\tau_g,\tau_t})}^{(\alpha, \beta)}\} \mathbf{x}_j,
		\label{reel_eq}
	\end{equation}
	where $\Re \{ .\}$ and $\Im \{. \}$ are the real and imaginary parts of the complex argument, respectively. Setting $\frac{\partial f(\mathbf{x})}{\partial \mathbf{x}_j} = 0$ yields:
	\begin{equation}
		\Im \{\mathbf{y} \} = \mathbf{x}_j + \Im \{ {(\mathbf{L}_J^{\tau_g,\tau_t})}^{(\alpha, \beta)}\} \mathbf{x}_r + \Re \{{(\mathbf{L}_J^{\tau_g,\tau_t})}^{(\alpha, \beta)}\} \mathbf{x}_j.
		\label{im_eq}
	\end{equation}
	Now we multiply~\cref{im_eq} by $j$ and add to~\cref{reel_eq} to get:
	\begin{equation}
		\mathbf{y} = \mathbf{x} +{(\mathbf{L}_J^{\tau_g,\tau_t})}^{(\alpha, \beta)} \mathbf{x}_r + j {(\mathbf{L}_J^{\tau_g,\tau_t})}^{(\alpha, \beta)} \mathbf{x}_j = \mathbf{x} + {(\mathbf{L}_J^{\tau_g,\tau_t})}^{(\alpha, \beta)} \mathbf{x}.
	\end{equation}
	Using the mixed product property of the Kronecker product with the diagonalization of ${(\mathbf{L}_G)}_{\beta}$ and ${(\mathbf{L}_T)}_{\alpha}$, we can write:
	\begin{align}
		{(\mathbf{L}_J^{\tau_g,\tau_t})}^{(\alpha, \beta)} & = {(\tau_t \mathbf{L}_T)}_{\alpha} \otimes \mathbf{I}_G + \mathbf{I}_T \otimes {(\tau_g \mathbf{L}_G)}_{\beta} \nonumber                                                                                \\ &= (\mathbf{U}_T^{\alpha} (\tau_t \mathbf{\Lambda}_T^{\alpha}) {(\mathbf{U}_T^{\alpha})}^H) \otimes \mathbf{I}_G \nonumber \\ &+ \mathbf{I}_T \otimes (\mathbf{U}_G^{\beta} (\tau_g
		\mathbf{\Lambda}_G^{\beta}) {(\mathbf{U}_G^{\beta})}^H) \nonumber                                                                                                                                                                                            \\
		                                                   & = (\mathbf{U}_T^{\alpha} \otimes \mathbf{U}_G^{\beta}) (\tau_t \mathbf{\Lambda}_T^{\alpha} \oplus \tau_g \mathbf{\Lambda}_G^{\beta}) {(\mathbf{U}_T^{\alpha} \otimes \mathbf{U}_G^{\beta})}^H \nonumber \\ &= \mathbf{U}_J^{(\alpha,\beta)} {(\mathbf{\Lambda}_J^{\tau_g, \tau_t})}^{(\alpha,\beta)} {(\mathbf{U}_J^{(\alpha,\beta)})}^H,
		\label{modified_decomposition}
	\end{align}
	where ${(\mathbf{\Lambda}_J^{\tau_g, \tau_t})}^{(\alpha,\beta)}\triangleq  (\tau_t \mathbf{\Lambda}_T^{\alpha} \oplus \tau_g \mathbf{\Lambda}_G^{\beta})$.
	Hence, JFRT also diagonalizes the regularized joint fractional Laplacian. Then, using~\cref{modified_decomposition}, we have the optimal $\mathbf{\hat{x}}$ in terms of the diagonalized form of the optimal filter matrix applied to $\mathbf{y}$ as:
	\begin{align}
		\mathbf{\hat{x}} & = {(\mathbf{I} + {(\mathbf{L}_J^{\tau_g,\tau_t})}^{(\alpha, \beta)})}^{-1} \mathbf{y} \nonumber                                                                          \\
		                 & = \mathbf{U}_J^{(\alpha,\beta)} {(\mathbf{I} + {(\mathbf{\Lambda}_J^{\tau_g,\tau_t})}^{(\alpha, \beta)})}^{-1} {(\mathbf{U}_J^{(\alpha,\beta)})}^H \mathbf{y}, \nonumber
	\end{align}
	from which we lastly obtain the optimal filter coefficients in the joint fractional time-vertex Fourier domains as:
	\begin{equation*}
		h^{(\alpha,\beta)}_{m,n} = \frac{1}{1+\tau_g \lambda_m^{\beta} + \tau_t \omega_n^{\alpha}}.
	\end{equation*}
\end{proof}

\section{Experiments and Results}\label{sect_exp}
For numerical experiments, we first provide applications of JFRT, such as denoising and clustering. These experiments demonstrate the superiority of the JFRT over JFT and other graph filtering approaches by providing better results with the same asymptotic computational complexity as JFT\@. Even though the literature on filtering and denoising of static graph signals are abundant, the filtering and denoising for time-varying graph signals on underlying static graphs are mainly achieved by JFT-based filtering~\cite{grassi18timevertex}, ARMA graph filtering~\cite{isufi17armafilter}, median filtering~\cite{tay21timevaryingdenoising}, and GNN-based~\cite{rey22untrainedgnn} methods. We inherently compare the proposed JFRT-based approaches with the JFT-based ones and provide comprehensive comparisons with the above state-of-the-art methods. We also provide the median filter~\cite{tay21timevaryingdenoising}, ARMA graph filter~\cite{isufi17armafilter}, untrained GNN~\cite{rey22untrainedgnn} and TimeGNN~\cite{castro23timegnn} implementation details in~\ref{sec:app:median},~\ref{sec:app:arma}~\ref{sec:untrained} and~\ref{sec:timegnn}, respectively.

\subsection{General Denoising Experiments}\label{sec:exp:denoising}
To illustrate fractional joint time-vertex filtering proposed in Section~\ref{sec:jtv_frac_filter_theory}, we design and perform denoising experiments on (1) synthetic data, and (2) real-world data. For the denoising task, we design a joint filter, a separable ideal low-pass filter in both fractional graph spectral and fractional frequency domains. We implement the separable joint low-pass filter based on the derivation in Section~\ref{sec:seperable_frac_jtv_filter_theory} by ordering the graph frequencies based on the GFT definition, as in Section~\ref{sec:graph_freq_ordering}.

We select the fractional graph spectral and frequency domain filters as ideal low-pass filters, ideal in the following sense: The graph ideal low-pass filter \(\mathbf{H}_G\) is defined as a diagonal matrix, where the last \(c\) and first \(N-c\) diagonal entries are set to \(0\) and \(1\), respectively. The time ideal low-pass filter \(\mathbf{H}_T\) is defined as a diagonal matrix, whose first and last \(n\) entries are set to \(1\), and the remaining \(T-2n\) entries are set to \(0\), where matrix representations are as follows:
\begin{equation}\label{eq:low_exp:graph_filters}
	\mathbf{H}_G^\beta =
	\begin{bmatrix}
		\mathbf{I}_{N - c} &                \\
		                   & \mathbf{0}_{c}
	\end{bmatrix},\quad
	{(\mathbf{H}_T^\alpha)}^T =
	\begin{bmatrix}
		\mathbf{I}_{n} &                     &              \\
		               & \mathbf{0}_{T - 2n} &              \\
		               &                     & \mathbf{I}_n \\
	\end{bmatrix}.
\end{equation}
Then, the problem can be formalized for a joint time-vertex signal \(\mathbf{X}\in\mathbb{R}^{N\times T}\), where each vertex has a delayed version of the original chirp signal. For a noise matrix \(\mathbf{N}\in\mathbb{R}^{N\times T}\), whose entries are i.i.d.\ from zero mean Gaussian, and we have \(\mathbf{Y} = \mathbf{X} + \mathbf{N}\). Then, we obtain the estimate of the original joint time-vertex signal \(\hat{\mathbf{X}}\) as:
\begin{equation}
	\hat{\mathbf{X}} = \mathbf{F}_G^{-\beta}\mathbf{H}_G^{\beta}\mathbf{F}_G^{\beta}\,\mathbf{Y}\,{\left(\mathbf{F}^{\alpha}\right)}^T{\left(\mathbf{H}_T^{\alpha}\right)}^T{\left(\mathbf{F}^{-\alpha}\right)}^T.
\end{equation}
Based on this filtering, we compute the noise error \(e_n\) and the estimation error \(e_e\) as follows in the form of percentage root mean squared error (RMSE (\%)):
\begin{equation}
	e_{n} \triangleq 100\times\frac{{\lVert\mathbf{X} - \mathbf{Y}\rVert}_F}{{\lVert\mathbf{X}\rVert}_F},\quad
	e_{e} \triangleq 100\times\frac{{\lVert\mathbf{X} - \hat{\mathbf{X}}\rVert}_F}{{\lVert\mathbf{X}\rVert}_F}.
\end{equation}
\subsubsection{Synthetic Data}
We model a physical scenario where we assume that we have a sensor network whose sensors measure the same signal in a noisy channel and with a time delay according to their positions. Delay for each vertex is selected proportional to the sensor locations to generate low-frequency graph signals. The base signal is selected as the chirp (sweep) signal to include various frequencies. The additive noise is selected to be i.i.d.\ zero mean Gaussian. Finally, the underlying graph is selected as the \emph{David Sensor Network} from the \emph{GSP toolbox}~\cite{perraudin14gspbox}, with \(N=64\) vertices. In this sensor network, the sensors, i.e., the vertices, are placed randomly in the unit square, where each sensor is connected to other sensors in the fixed radius of itself, where fixed radius selection also enforces the graph to be undirected. The underlying toolbox has utilized the thresholded Gaussian kernel to generate the edge weights, and finally, we ensure that the generated graph is connected. This experiment uses adjacency and Laplacian-based GFT definitions. The adjacency matrix is row normalized with its degree matrix \(\mathbf{D}\) such that \(\mathbf{A}^\prime = \mathbf{D}^{-1}\mathbf{A}\) to prevent scaling.

\begin{figure}[ht]
	\centering{}
	\begin{subfigure}{0.32\linewidth}
		\centering{}
		\includegraphics[width=\linewidth, height=0.17\textheight, keepaspectratio]{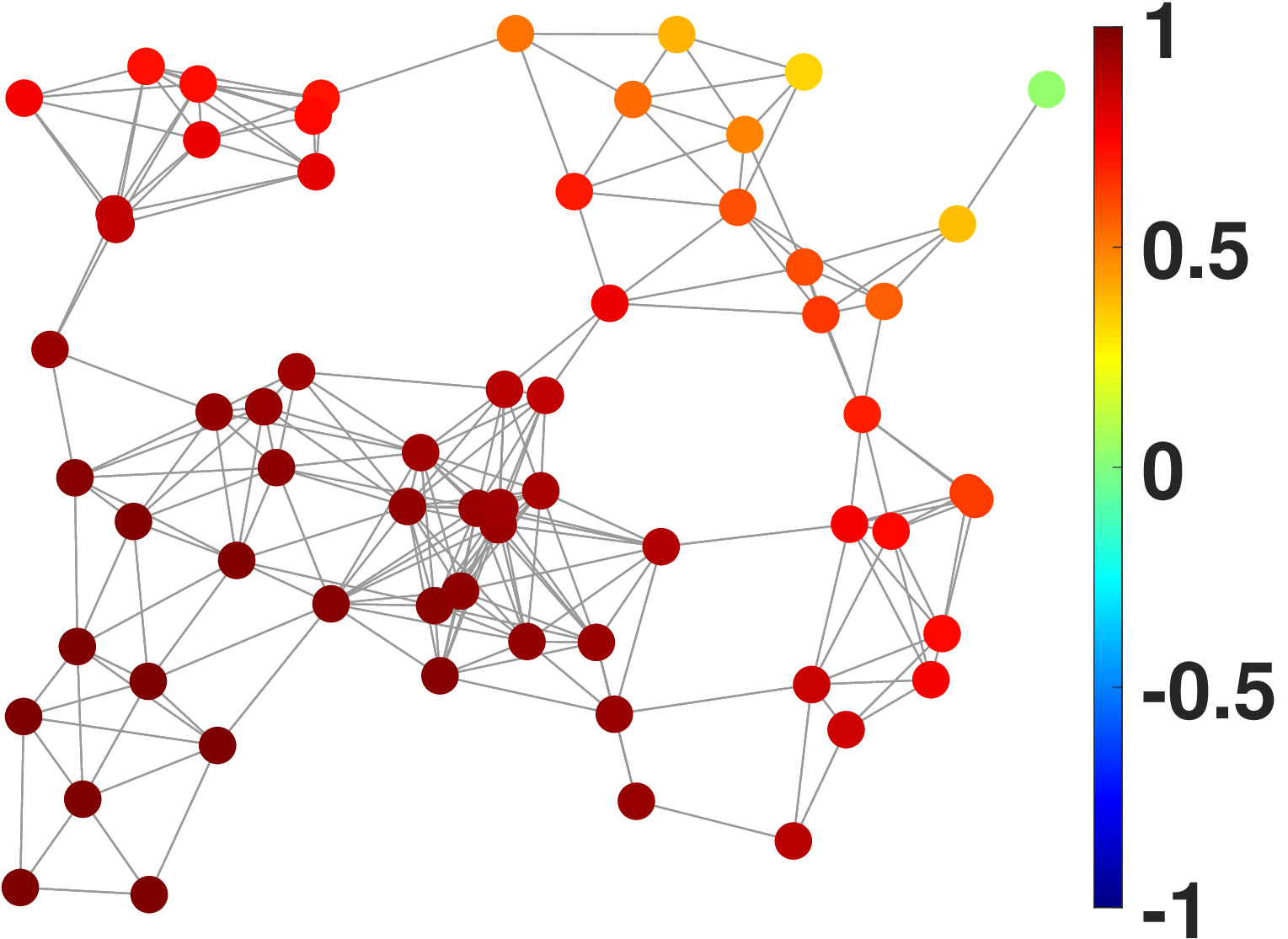}
	\end{subfigure}
	\begin{subfigure}{0.32\linewidth}
		\centering{}
		\includegraphics[width=\linewidth, height=0.17\textheight, keepaspectratio]{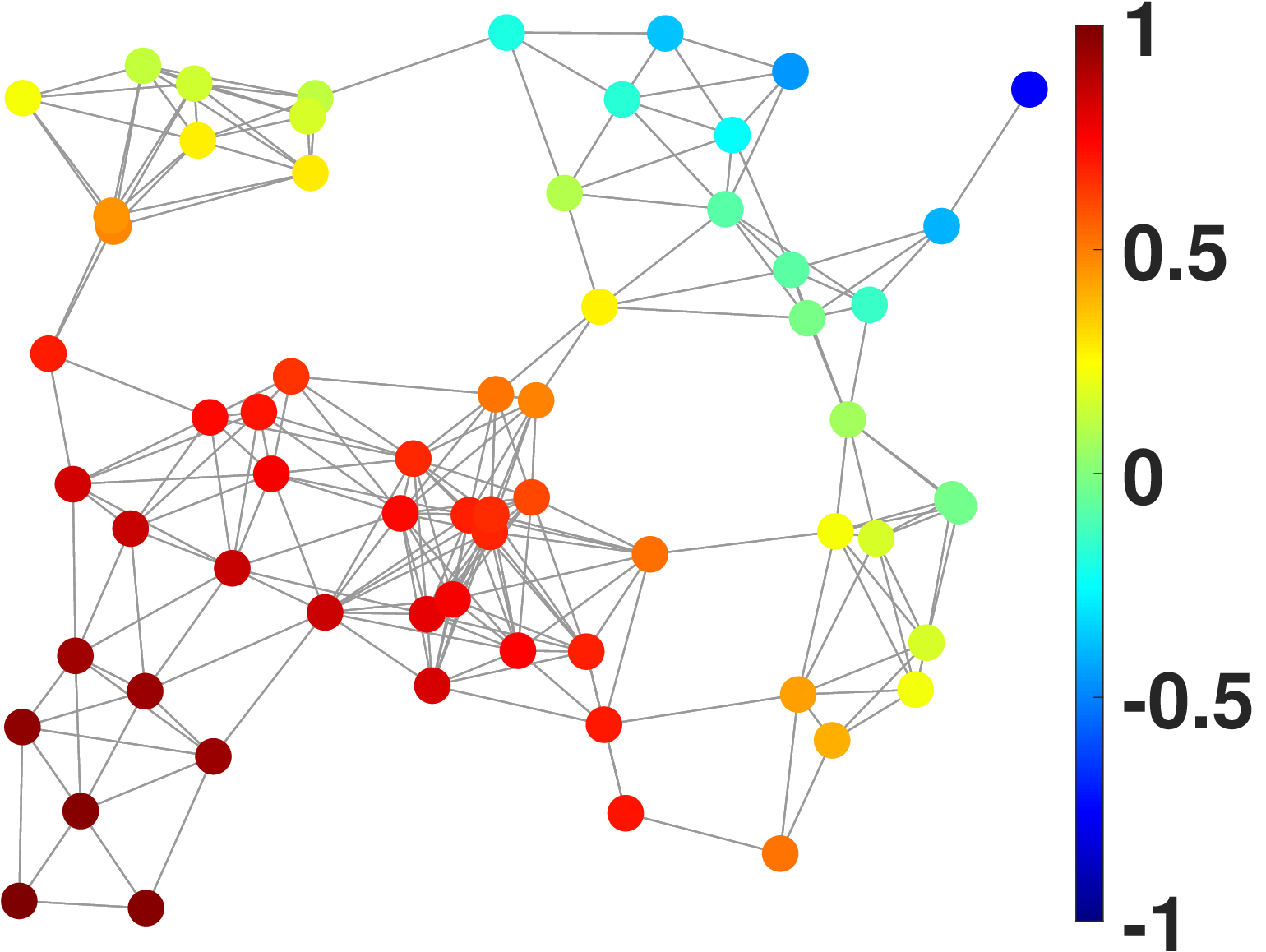}
	\end{subfigure}
	\begin{subfigure}{0.32\linewidth}
		\centering{}
		\includegraphics[width=\linewidth, height=0.17\textheight, keepaspectratio]{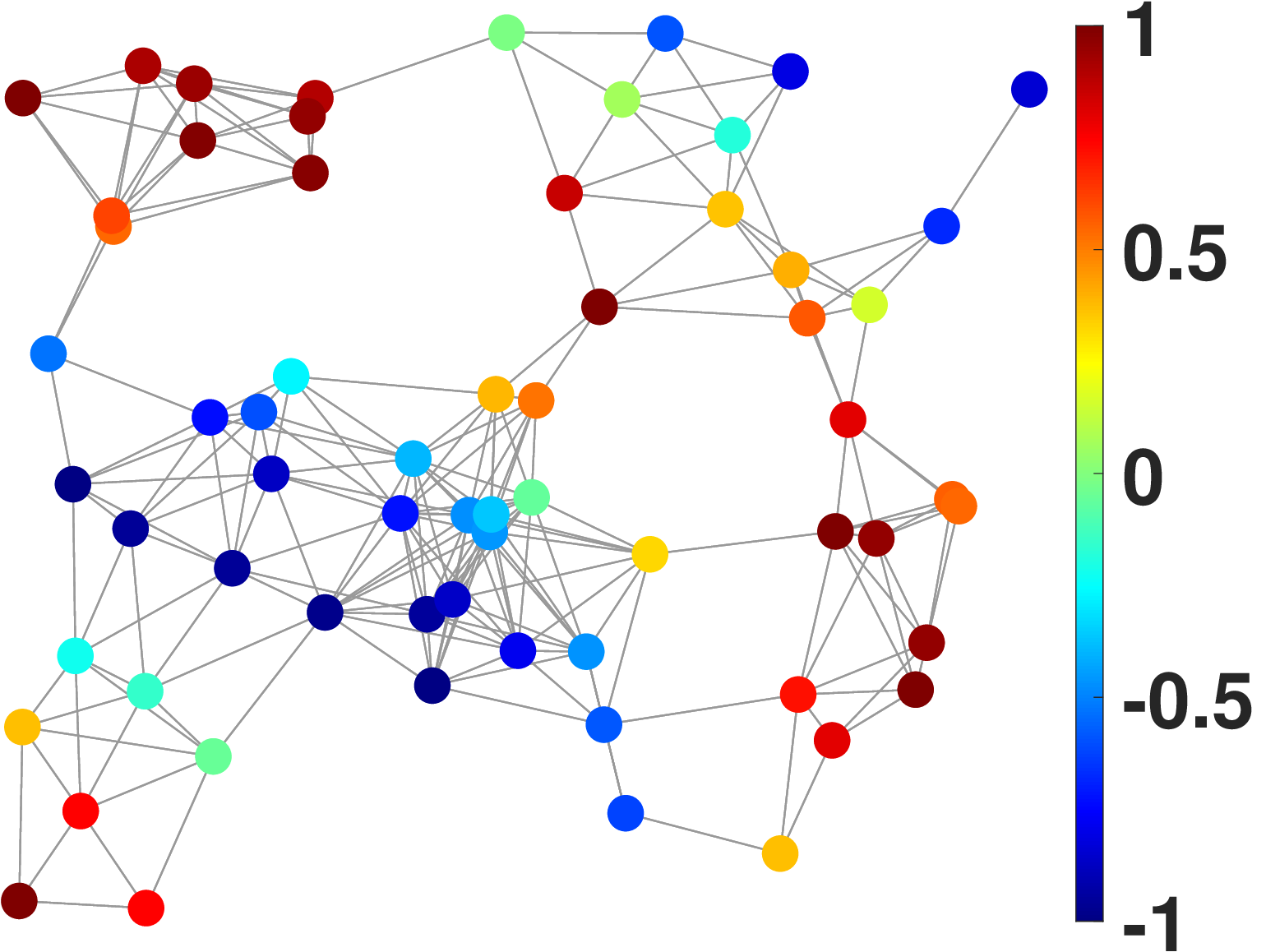}
	\end{subfigure}
	\caption{Graph signals for the delay parameter \(d=35\), and at time instances, from left-to-right, \(t=0, 0.009, 0.099\) secs., which correspond to the \(1^{\text{st}}\), \({10}^{\text{th}}\) and \({100}^{\text{th}}\) columns of the joint time-vertex signal \(\mathbf{X}\), where \(f_s = 1\,\si{\kilo\hertz}\).}~\label{fig:filter:graph_signal}
\end{figure}

\begin{figure}[ht]
	\centering{}
	\begin{subfigure}[t]{0.32\linewidth}
		\centering{}
		\includegraphics[width=\linewidth, height=0.17\textheight, keepaspectratio]{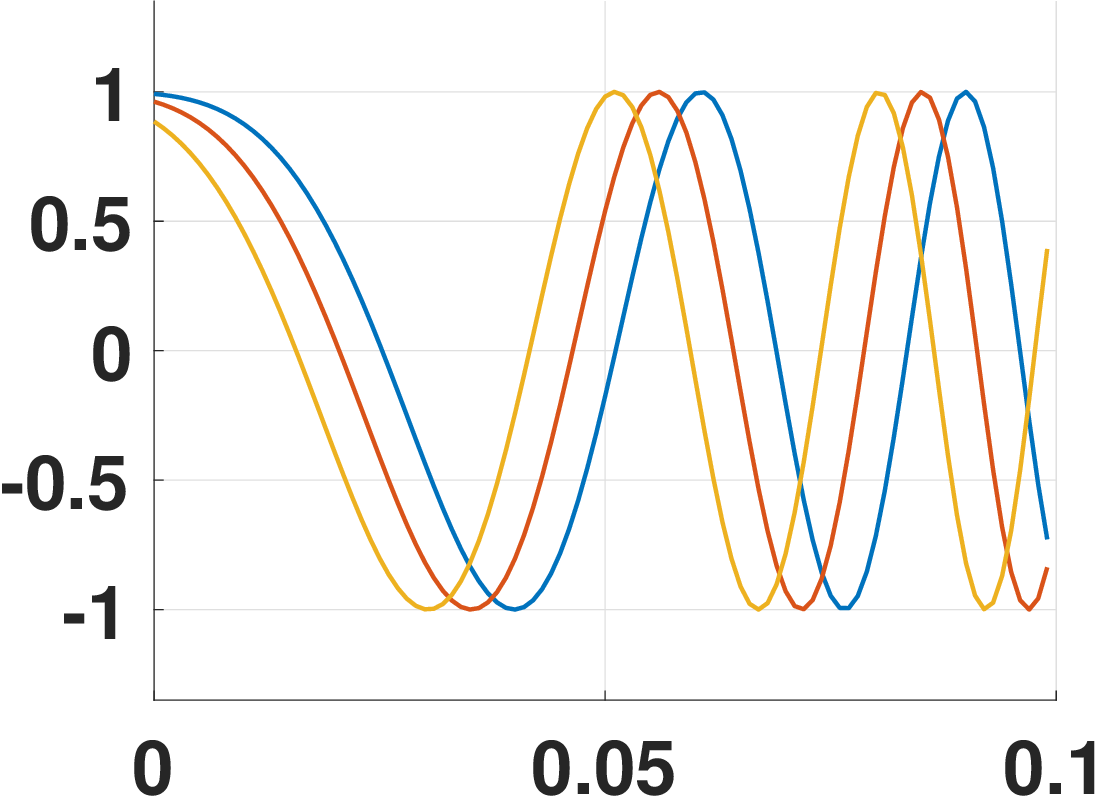}
		\caption{Original}\label{fig:filter:time_signal:orig}
	\end{subfigure}
	\begin{subfigure}[t]{0.32\linewidth}
		\centering{}
		\includegraphics[width=\linewidth,height=0.17\textheight, keepaspectratio]{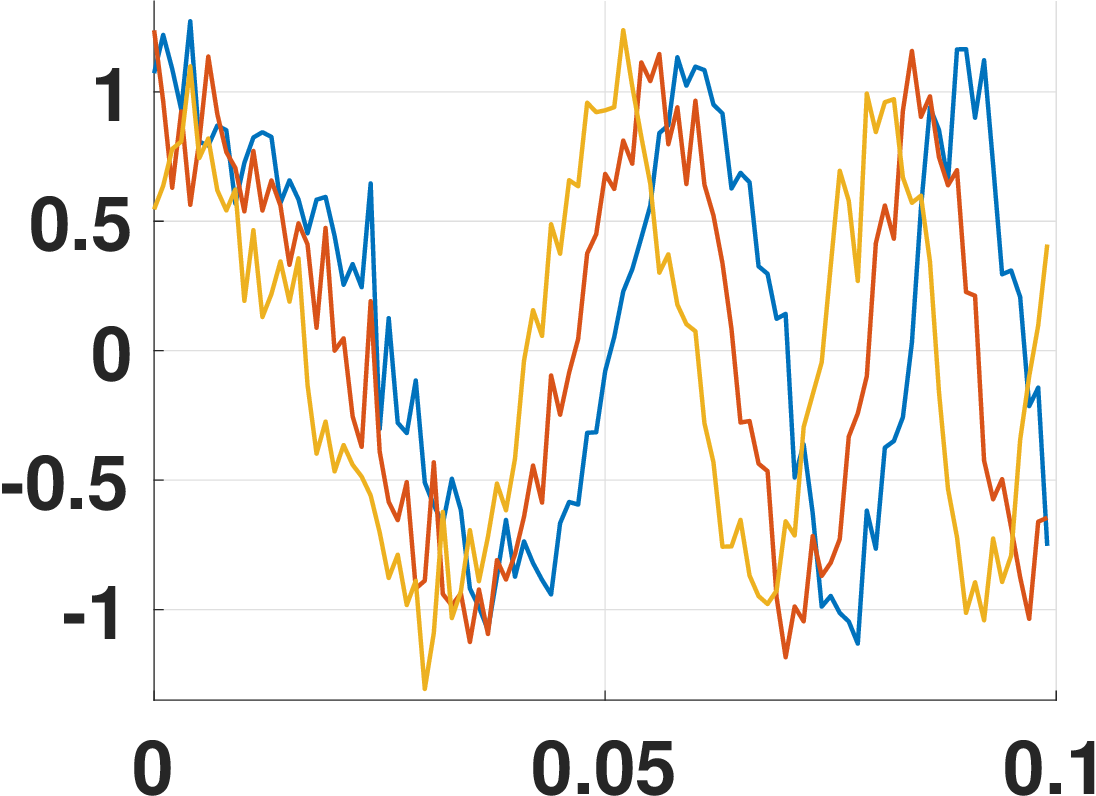}
		\caption{Noisy}\label{fig:filter:time_signal:noisy}
	\end{subfigure}
	\begin{subfigure}[t]{0.32\linewidth}
		\centering{}
		\includegraphics[width=\linewidth,height=0.17\textheight, keepaspectratio]{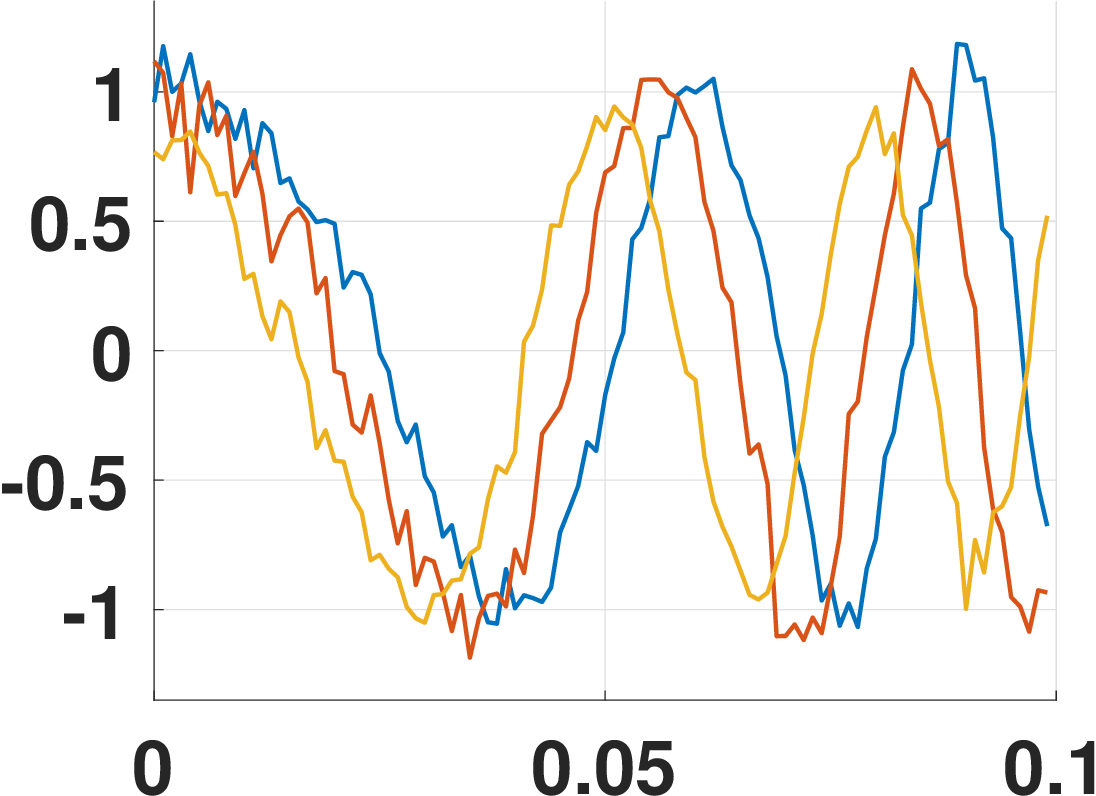}
		\caption{Filtered}\label{fig:filter:time_signal:filtered}
	\end{subfigure}
	\caption{Time series signals for the first three vertices in the sensor network. The original version is obtained with the \(d=30\), the noisy version with \(\sigma=0.15\), and the filtered version with Adjacency method and \(\alpha = 1.34, \beta=1.01\), and \(c=35\). Error reduces to \(16.42\% \) from \(21.36\% \) after filtering.}~\label{fig:filter:time_signal}
\end{figure}

To generate the delayed chirp signals, we employ the following procedure: First, we generate the baseline chirp signal with the lowest frequency \(f_{\min} = 0\,\si{\hertz}\), the highest frequency \(f_{\max} = 400\,\si{\hertz}\) and sampling frequency \(f_s = 1\,\si{\kilo\hertz}\), where the duration of the chirp is \(T_{\text{dur}} = 1\,\si{\second}\). First \(100\) samples are taken from the chirp signal, i.e., \(T_{\text{dur}}^\prime = 0.1\,\si{\second}\) and \(T = 100\) for joint time-vertex signal. After generating the baseline chirp signal, we introduce the delays for different vertices. We exploit the unit square nature of the underlying graph to apply similar delays to vertices that are close to each other. The location vector norms are calculated for each vertex and normalized to the maximum norm, which we refer to as \({\lVert\mathbf{r}_i\rVert}_2\) for the \(i\)th vertex. Then, we define the delay for the \(i\)th vertex as \(t_{\text{delay}}^{(i)} = \frac{{\lVert\mathbf{r}_i\rVert}_2}{f_s}\times d\), where \(d\) is the delay multiplier, and we present results for \(d\in \{25,30,35,40,45,50\} \). If we denote the \(i\)th row of the joint time-vertex signal, \(\mathbf{X}\), as \(\mathbf{x}_i^T\), then \(\mathbf{x}_i^T = \texttt{chirp}(t + t_{\text{delay}}^{(i)})\), where \(\texttt{chirp}(t)\) generates the baseline chirp signal. The resulting joint time-vertex signal is partially presented in graph and time domains in Figs.~\ref{fig:filter:graph_signal} and~\ref{fig:filter:time_signal}, respectively.
\begin{figure}[ht]
	\centering{}
	\begin{subfigure}{0.49\linewidth}
		\centering{}
		\includegraphics[width=\linewidth, height=0.2\textheight, keepaspectratio]{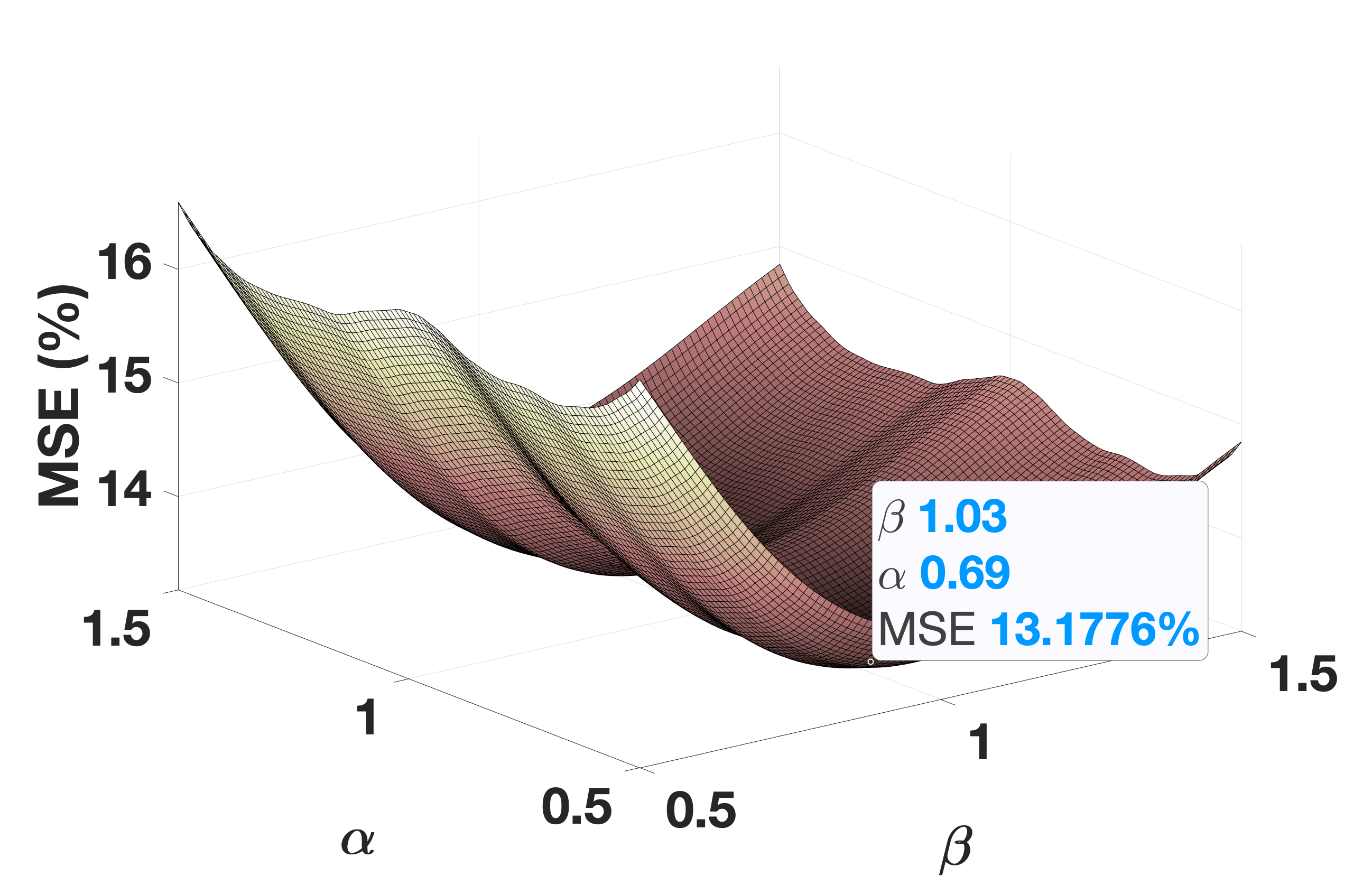}
	\end{subfigure}
	\begin{subfigure}{0.49\linewidth}
		\centering{}
		\includegraphics[width=\linewidth, height=0.2\textheight, keepaspectratio]{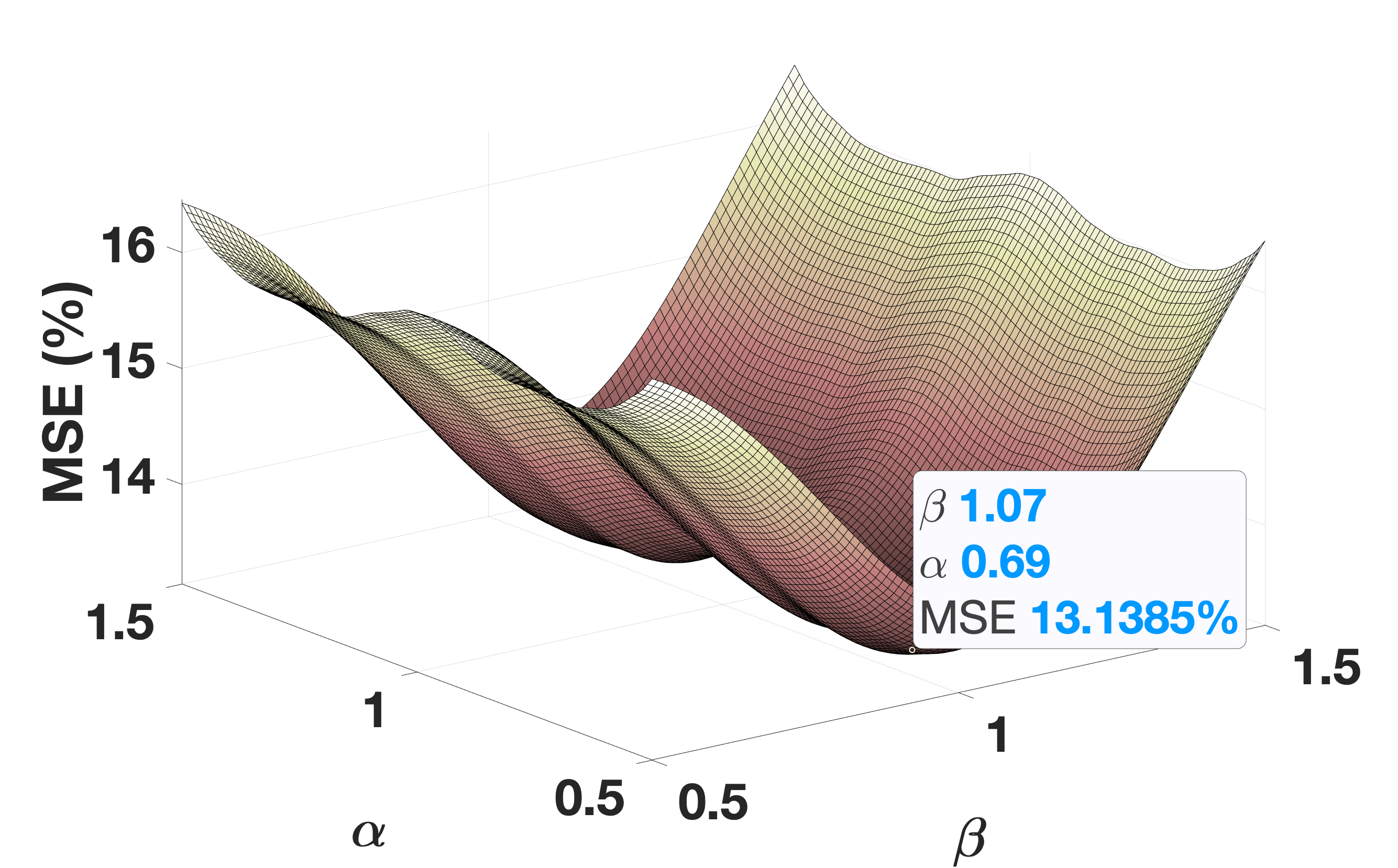}
	\end{subfigure}
	\caption{Error surfaces for both the adjacency (left) and Laplacian (right) based joint time-vertex fractional filtering methods, for \(\sigma=0.1,d=50, c=1\). Best \(\alpha,\beta \) pairs are obtained according to the defined error metric and found to be \({(\alpha,\beta)}_{\text{adj}}=(0.69,1.03)\) and \({(\alpha,\beta)}_{\text{lap}}=(0.69,1.07)\).}~\label{fig:filter:error_surface}
\end{figure}

In Fig.~\ref{fig:filter:graph_signal}, three-time instances are selected to demonstrate the graph signals of the joint time-vertex signal \(\mathbf{X}\). The selected time instances correspond to the \(1^{\text{st}}\), \({10}^{\text{th}}\) and \({100}^{\text{th}}\) columns of \(\mathbf{X}\), where the delay parameter is \(d=35\), demonstrating the low-frequency nature of the graph signals, especially in the lower time instances. In Fig.~\ref{fig:filter:time_signal:orig}, the time-series signals at the first three vertices are shown, which correspond to the first three rows of \(\mathbf{X}\), with delay parameter \(d=30\). We note that the vertex index and location are not correlated with the underlying graph. We then introduce the noise matrix \(\mathbf{N}\), whose entries are i.i.d.\ from zero mean Gaussian, with different standard deviations \(\sigma\in \{0.1,0.15,0.2\} \). The resulting noisy joint time-vertex signal \(\mathbf{Y}\) is shown in Fig.~\ref{fig:filter:time_signal:noisy} for \(\sigma=0.15\) case.

Finally, we apply the JFRT filtering by selecting the cut-off frequency \(f_c = 450\,\si{\hertz}\) for the time domain, and by fixing \(d\) and \(\sigma \) values, to find the optimal triplet of \((\alpha,\beta,c)\) for both adjacency and Laplacian based methods, where \(\alpha\in\interval{0.5}{1.5}\), \(\beta\in\interval{0.5}{1.5}\) and \(c\in \{0,1,\dots,(N-1)=63\} \). We also indicate selecting a cut-off frequency \(f_c\) determines \(n\) for \(\mathbf{H}_T\) as \(n = \left\lfloor T f_c/f_s \right\rfloor \) for sampling frequency \(f_s\) and cut-off frequency \(f_c\) such that \(f_c < f_s/2\). For \(f_c \geq f_s/2\), the time domain ideal low-pass filter reduces to \(\mathbf{I}_T\). These triplets are obtained for each standard deviation value, delay parameter, and GFT method, where the best-performing settings are tabulated in~\ref{sec:app:experiments}. Example of the \((\alpha,\beta)\) error surfaces for configuration of \(\sigma=0.1,d=50\) is also provided in Fig.~\ref{fig:filter:error_surface}, for \(c=1\). The filtered version of the noisy signal is provided in Fig.~\ref{fig:filter:time_signal:filtered}, with \((\alpha,\beta,c) = (1.34,1.01,35)\), which is the optimal setting for the given configuration of \(\sigma=0.15\) and \(d=30\).

\subsubsection{Real-World Data}\label{sec:denoise:real}
We also conduct the same experiment on real-world datasets of Sea Surface Temperature (SST) and COVID19-USA, where~\cite{giraldo22sobolev} provides the pre-processed versions. The SST dataset consists of the monthly captured sea surface temperatures provided by the NOAA Physical Sciences Laboratory.  We use the subset of first \(N = 100\) vertices on the Pacific Ocean within a time frame of first \(T = 120\) months as in~\cite{giraldo22sobolev,qiu17timevar,tay21timevaryingdenoising}. Johns Hopkins University provides the COVID19-USA dataset using the version provided by~\cite{giraldo22sobolev}. The dataset consists of the daily confirmed cases of COVID-19 in the $3,232$ localities in the USA between January 22, 2020, and November 18, 2020. We use the first \(N = 100\) vertices and \(T = 302\) days. The SST dataset can be seen as a discretization of a manifold, whereas the COVID19-USA dataset is naturally discrete. We use \textit{gspbox}~\cite{perraudin14gspbox} to generate $k$-nearest neighbors ($k$-NN) graphs for both datasets as in~\cite{giraldo22sobolev,qiu17timevar,tay21timevaryingdenoising}. We experiment with several \(k\in \{2,5,10\} \) values, where $k$-NN generation sigma selected as \(\sigma_{\text{$k$-NN}}=1,000\) to have non-sparse graphs for all selected \(k\) values. We also normalize the joint time-vertex graph signals to the interval \(\interval{0}{1}\).

\subsubsection{Results and Comparison}
We compare our results with the denoising methods of JFT~\cite{grassi18timevertex}, ARMA graph filter~\cite{isufi17armafilter}, median filter~\cite{tay21timevaryingdenoising}, TimeGNN~\cite{castro23timegnn} and GCN~\cite{kipf17gcn}. For ARMA graph filters, we use the parallel implementation of \(\text{ARMA}_K\) filters, as suggested in~\cite{isufi17armafilter}, with \(K\in \{3,4,5\} \). For time-varying median filters, we use both \(\mathcal{M}_1\) and \(\mathcal{M}_2\) filtering approaches described in~\cite{tay21timevaryingdenoising}, which we refer as \(\text{Median}_1\) and \(\text{Median}_2\), respectively. For GNN-based denoising, we employ untrained GNN framework~\cite{rey22untrainedgnn}. The denoising experiment results for synthetic, SST, and COVID19-USA datasets are provided in Tables~\ref{tab:filter:comparison},~\ref{tab:denoise:sst} and~\ref{tab:denoise:covid}, respectively. We provide GNN-based method comparisons only for real-world datasets. JFRT-based denoising outperforms other methods in almost every setting. We provide the best-performing settings in~\ref{sec:app:experiments}.
\begin{table}[ht]
	\centering
	\caption{Estimation error (RMSE (\%) $\downarrow$) comparison for the denoising experiment on synthetic data, where \(d\) is the delay multiplier, \(\sigma \) is Gaussian noise standard deviation, and \((e_n,e_e)\) are the noise and estimation errors, respectively.}\label{tab:filter:comparison}
	\resizebox{\linewidth}{!}{
		\begin{tabular}{@{}llrrrrrrrrrrrrrrrrrr@{}}
			\toprule
			\multirow{3}{*}{\textbf{Method}}               & \(\boldsymbol{d}\)       & \multicolumn{3}{c}{25} & \multicolumn{3}{c}{30} & \multicolumn{3}{c}{35} & \multicolumn{3}{c}{40} & \multicolumn{3}{c}{45} & \multicolumn{3}{c}{50}                                                                                                                                                                                                             \\ \cmidrule(l){3-5}\cmidrule(l){6-8}\cmidrule(l){9-11}\cmidrule(l){9-11}\cmidrule(l){12-14}\cmidrule(l){15-17}\cmidrule(l){18-20}
			                                               & \(\boldsymbol{\sigma }\) & 0.10                   & 0.15                   & 0.20                   & 0.10                   & 0.15                   & 0.20                   & 0.10           & 0.15           & 0.20           & 0.10           & 0.15           & 0.20           & 0.10           & 0.15           & 0.20           & 0.10           & 0.15           & 0.20           \\
			                                               & \(\boldsymbol{e_n}\)     & 14.13                  & 21.19                  & 28.25                  & 14.24                  & 21.36                  & 28.48                  & 14.30          & 21.45          & 28.59          & 14.35          & 21.53          & 28.69          & 14.33          & 21.49          & 28.65          & 14.25          & 21.37          & 28.49          \\ \midrule
			\(\text{Median}_1\)                            &                          & 14.48                  & 15.61                  & \textbf{17.01}         & 16.88                  & 18.01                  & \textbf{19.39}         & 19.59          & 20.64          & 21.92          & 22.58          & 23.76          & 25.09          & 25.62          & 26.81          & 28.13          & 28.57          & 29.61          & 30.94          \\
			\(\text{Median}_2\)                            &                          & 16.54                  & 17.10                  & 17.76                  & 20.67                  & 21.00                  & 21.47                  & 25.31          & 25.48          & 25.82          & 30.25          & 30.40          & 30.62          & 35.34          & 35.41          & 35.57          & 40.63          & 40.58          & 40.70          \\
			\(\text{ARMA}_3\)                              &                          & 15.50                  & 21.17                  & 27.17                  & 15.52                  & 21.24                  & 27.29                  & 15.51          & 21.25          & 27.32          & 15.53          & 21.28          & 27.38          & 15.53          & 21.25          & 27.33          & 15.48          & 21.15          & 27.19          \\
			\(\text{ARMA}_4\)                              &                          & 13.50                  & 20.01                  & 26.58                  & 13.61                  & 20.17                  & 26.78                  & 13.71          & 20.28          & 26.90          & 13.83          & 20.40          & 27.04          & 13.88          & 20.41          & 27.03          & 13.86          & 20.32          & 26.88          \\
			\(\text{ARMA}_5\)                              &                          & 13.70                  & 19.58                  & 25.63                  & 13.87                  & 19.75                  & 25.82                  & 14.03          & 19.89          & 25.97          & 14.26          & 20.09          & 26.16          & 14.47          & 20.22          & 26.24          & 14.61          & 20.25          & 26.20          \\
			\(\text{JFT}_{\text{Adj}}\)                    &                          & 11.91                  & 15.74                  & 18.78                  & 13.07                  & 16.95                  & 20.95                  & 13.46          & 18.27          & 22.17          & 13.65          & 19.53          & 23.76          & 13.62          & 19.78          & 25.28          & 13.57          & 19.99          & 25.77          \\
			\(\text{JFT}_{\text{Lap}}\)                    &                          & 12.18                  & 16.47                  & 18.94                  & 12.74                  & 17.39                  & 21.05                  & 13.22          & 18.21          & 22.77          & 13.50          & 19.07          & 23.69          & 13.54          & 19.78          & 24.87          & 13.59          & 19.97          & 25.89          \\
			\(\text{\textbf{JFRT}}_{\text{\textbf{Adj}}}\) &                          & \textbf{11.37}         & \textbf{15.48}         & 18.27                  & 12.54                  & \textbf{16.42}         & 20.30                  & 12.98          & \textbf{17.99} & \textbf{21.57} & 13.23          & 19.21          & 23.15          & 13.22          & 19.56          & 24.67          & 13.17          & 19.56          & \textbf{25.17} \\
			\(\text{\textbf{JFRT}}_{\text{\textbf{Lap}}}\) &                          & 11.60                  & 16.02                  & 18.31                  & \textbf{12.20}         & 16.95                  & 20.55                  & \textbf{12.75} & 18.08          & 22.15          & \textbf{13.05} & \textbf{18.63} & \textbf{23.12} & \textbf{13.12} & \textbf{19.54} & \textbf{24.23} & \textbf{13.13} & \textbf{19.30} & 25.28          \\ \bottomrule
		\end{tabular}
	}
\end{table}
\begin{table}[ht]
	\footnotesize
	\centering
	\caption{Denoising experiment (RMSE (\%) $\downarrow$) results on SST dataset, where \(N = 100,\,T = 120,\,\sigma_{\text{$k$-NN}} = 1000\).}\label{tab:denoise:sst}
	\begin{tabular}{@{}llrrrrrrrrr@{}}
		\toprule
		\multirow{3}{*}{\textbf{Method}}               &                          & \multicolumn{3}{c}{\(k = 2\)} & \multicolumn{3}{c}{\(k = 5\)} & \multicolumn{3}{c}{\(k = 10\)}                                                                                                  \\ \cmidrule(l){3-5}\cmidrule(l){6-8}\cmidrule(l){9-11}
		                                               & \(\boldsymbol{\sigma }\) & \(0.10\)                      & \(0.15\)                      & \(0.20\)                       & \(0.10\)      & \(0.15\)      & \(0.20\)       & \(0.10\)      & \(0.15\)      & \(0.20\)      \\
		                                               & \(\boldsymbol{e_n}\)     & 14.57                         & 21.85                         & 29.13                          & 14.57         & 21.85         & 29.13          & 14.57         & 21.85         & 29.13         \\ \midrule
		\(\text{Median}_1\)                            &                          & 7.97                          & 11.51                         & 15.11                          & 7.72          & 10.29         & 12.95          & 9.05          & 10.65         & 12.40         \\
		\(\text{Median}_2\)                            &                          & 7.26                          & 9.57                          & 12.07                          & 7.86          & 9.16          & 10.66          & 9.77          & 10.35         & 11.14         \\
		\(\text{ARMA}_3\)                              &                          & 16.61                         & 22.06                         & 27.96                          & 15.58         & 21.48         & 27.70          & 14.54         & 21.21         & 28.00         \\
		\(\text{ARMA}_4\)                              &                          & 14.51                         & 21.76                         & 29.02                          & 15.58         & 21.48         & 27.70          & 14.54         & 21.21         & 28.00         \\
		\(\text{ARMA}_5\)                              &                          & 14.57                         & 21.85                         & 29.13                          & 13.99         & 20.15         & 26.47          & 14.45         & 20.05         & 25.93         \\
		GCN                                            &                          & 9.85                          & 9.93                          & 24.09                          & 8.21          & 8.73          & 32.99          & 39.35         & 16.66         & 39.32         \\
		TimeGNN                                        &                          & 96.64                         & 96.59                         & 96.54                          & 96.63         & 96.57         & 96.52          & 95.61         & 96.55         & 96.51         \\
		\(\text{JFT}_{\text{Adj}}\)                    &                          & 6.96                          & 8.92                          & 11.00                          & 6.79          & 8.69          & 10.61          & 7.16          & 8.23          & 9.53          \\
		\(\text{JFT}_{\text{Lap}}\)                    &                          & 6.42                          & 8.35                          & 10.46                          & 6.75          & 8.52          & 10.07          & 6.85          & 8.00          & 9.12          \\
		\(\text{\textbf{JFRT}}_{\text{\textbf{Adj}}}\) &                          & 6.94                          & 8.88                          & 10.95                          & 6.78          & 8.67          & 10.57          & 7.16          & 8.22          & 9.52          \\
		\(\text{\textbf{JFRT}}_{\text{\textbf{Lap}}}\) &                          & \textbf{6.41}                 & \textbf{8.31}                 & \textbf{10.40}                 & \textbf{6.73} & \textbf{8.50} & \textbf{10.04} & \textbf{6.84} & \textbf{7.99} & \textbf{9.11} \\  \bottomrule
	\end{tabular}
\end{table}
\begin{table}[ht]
	\footnotesize
	\centering
	\caption{Denoising experiment (RMSE (\%) $\downarrow$) results on \textit{COVID19-USA} dataset, where \(N = 100,\,T = 302,\,\sigma_{\text{$k$-NN}} = 1000\).}\label{tab:denoise:covid}
	\begin{tabular}{@{}llrrrrrrrrr@{}}
		\toprule
		\multirow{3}{*}{\textbf{Method}}               &                          & \multicolumn{3}{c}{\(k = 2\)} & \multicolumn{3}{c}{\(k = 5\)} & \multicolumn{3}{c}{\(k = 10\)}                                                                                                       \\ \cmidrule(l){3-5}\cmidrule(l){6-8}\cmidrule(l){9-11}
		                                               & \(\boldsymbol{\sigma }\) & \(0.010\)                     & \(0.015\)                     & \(0.020\)                      & \(0.010\)      & \(0.015\)      & \(0.020\)      & \(0.010\)      & \(0.015\)      & \(0.020\)      \\
		                                               & \(\boldsymbol{e_n}\)     & 13.80                         & 20.70                         & 27.60                          & 13.80          & 20.70          & 27.60          & 13.80          & 20.70          & 27.60          \\ \midrule
		\(\text{Median}_1\)                            &                          & 13.23                         & 16.01                         & \textbf{19.06}                 & 71.49          & 71.48          & 71.51          & 81.60          & 81.28          & 81.04          \\
		\(\text{Median}_2\)                            &                          & 63.55                         & 63.67                         & 63.90                          & 81.84          & 81.67          & 81.49          & 83.78          & 83.55          & 83.36          \\
		\(\text{ARMA}_3\)                              &                          & 44.25                         & 45.12                         & 46.31                          & 14.28          & 20.76          & 27.36          & 14.91          & 21.12          & 27.56          \\
		\(\text{ARMA}_4\)                              &                          & 44.25                         & 45.12                         & 46.31                          & 14.28          & 20.76          & 27.36          & 13.99          & 20.65          & 27.38          \\
		\(\text{ARMA}_5\)                              &                          & 26.75                         & 29.25                         & 32.42                          & 18.11          & 22.89          & 28.25          & 14.05          & 20.65          & 27.33          \\
		GCN                                            &                          & 45.30                         & 45.29                         & 45.31                          & 45.30          & 69.93          & 52.18          & 45.28          & 47.78          & 52.07          \\
		TimeGNN                                        &                          & 84.44                         & 82.74                         & 81.59                          & 82.44          & 80.73          & 79.76          & 81.21          & 79.64          & 78.83          \\
		\(\text{JFT}_{\text{Adj}}\)                    &                          & 11.09                         & 15.46                         & 20.03                          & 17.54          & 20.61          & 24.26          & 17.69          & 20.74          & 24.37          \\
		\(\text{JFT}_{\text{Lap}}\)                    &                          & 11.11                         & 15.47                         & 20.05                          & 12.95          & 16.87          & 21.17          & 12.57          & 16.58          & 20.93          \\
		\(\text{\textbf{JFRT}}_{\text{\textbf{Adj}}}\) &                          & 10.86                         & \textbf{15.31}                & 19.93                          & \textbf{10.94} & \textbf{15.40} & \textbf{20.04} & 15.53          & 18.95          & 22.89          \\
		\(\text{\textbf{JFRT}}_{\text{\textbf{Lap}}}\) &                          & \textbf{10.85}                & \textbf{15.31}                & 19.94                          & 11.44          & 15.76          & 20.31          & \textbf{11.44} & \textbf{15.73} & \textbf{20.25} \\
		\bottomrule
	\end{tabular}
\end{table}

\subsection{Tikhonov-Based Denoising Experiments}\label{sec:exp:tikhonov}
We performed denoising experiments on the Molene dataset~\cite{girault15transongraph} and the Weather station dataset (NOAA)~\cite{noaa20weather}. We implemented the denoising setting using JFRT as explained in Section~\ref{denoise_theory}. The open-access Molene dataset contains a graph of 37 weather stations across Northern France with 744 hourly temperature measurements~\cite{girault15transongraph}. Data in the Molene dataset are organized as a joint time-vertex signal on an undirected graph where the vertices are connected with their 5 nearest neighbors. The NOAA dataset contains the average temperatures measured by weather stations across Europe and the Middle East through 366 days of 2020. We have formed a 5 nearest-neighbor undirected graph using the distances between weather stations.

\begin{figure}[ht]
	\centering{}
	\begin{subfigure}{0.49\linewidth}
		\centering{}
		\includegraphics[width=\linewidth, height=0.2\textheight, keepaspectratio]{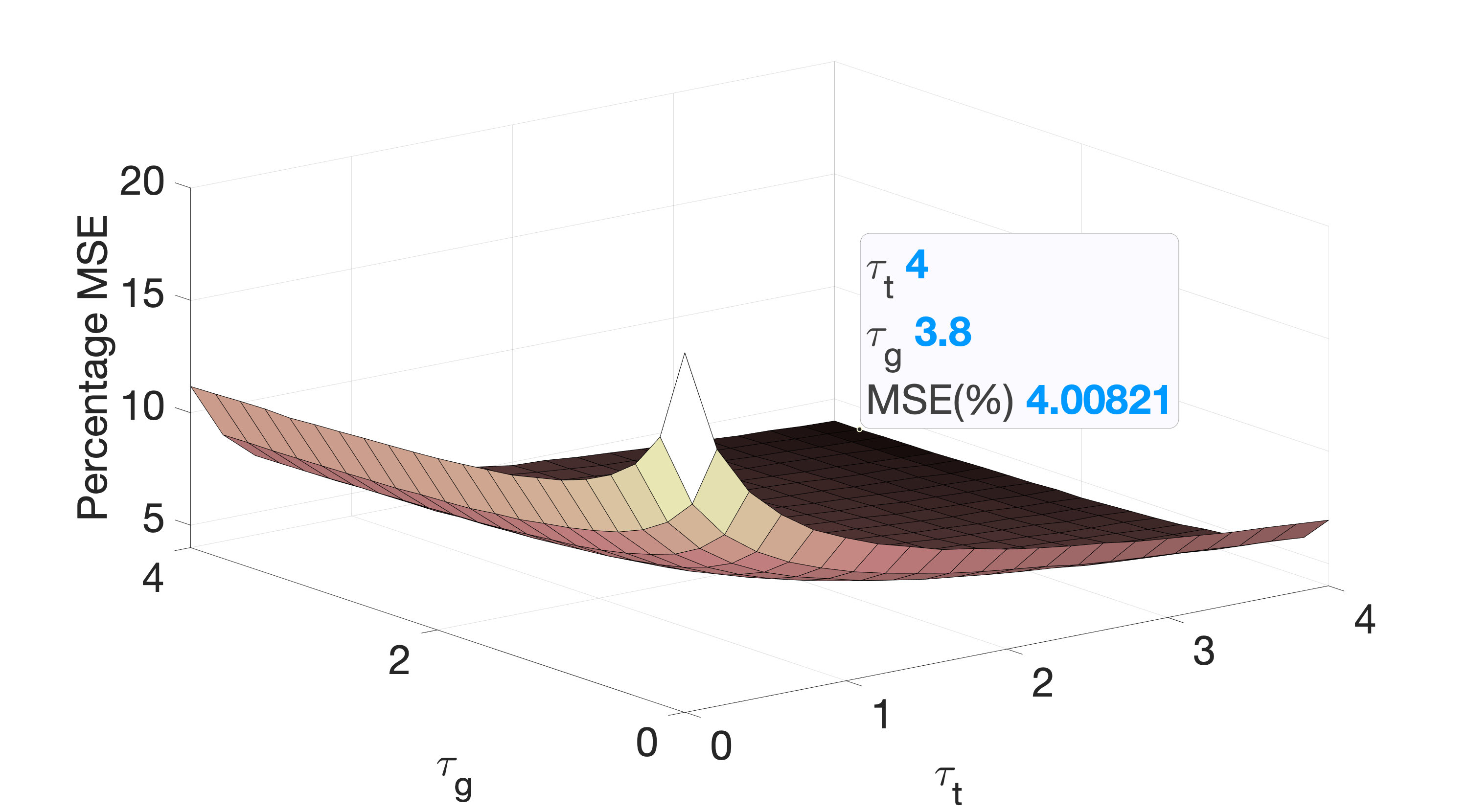}
	\end{subfigure}
	\begin{subfigure}{0.49\linewidth}
		\centering{}
		\includegraphics[width=\linewidth, height=0.2\textheight, keepaspectratio]{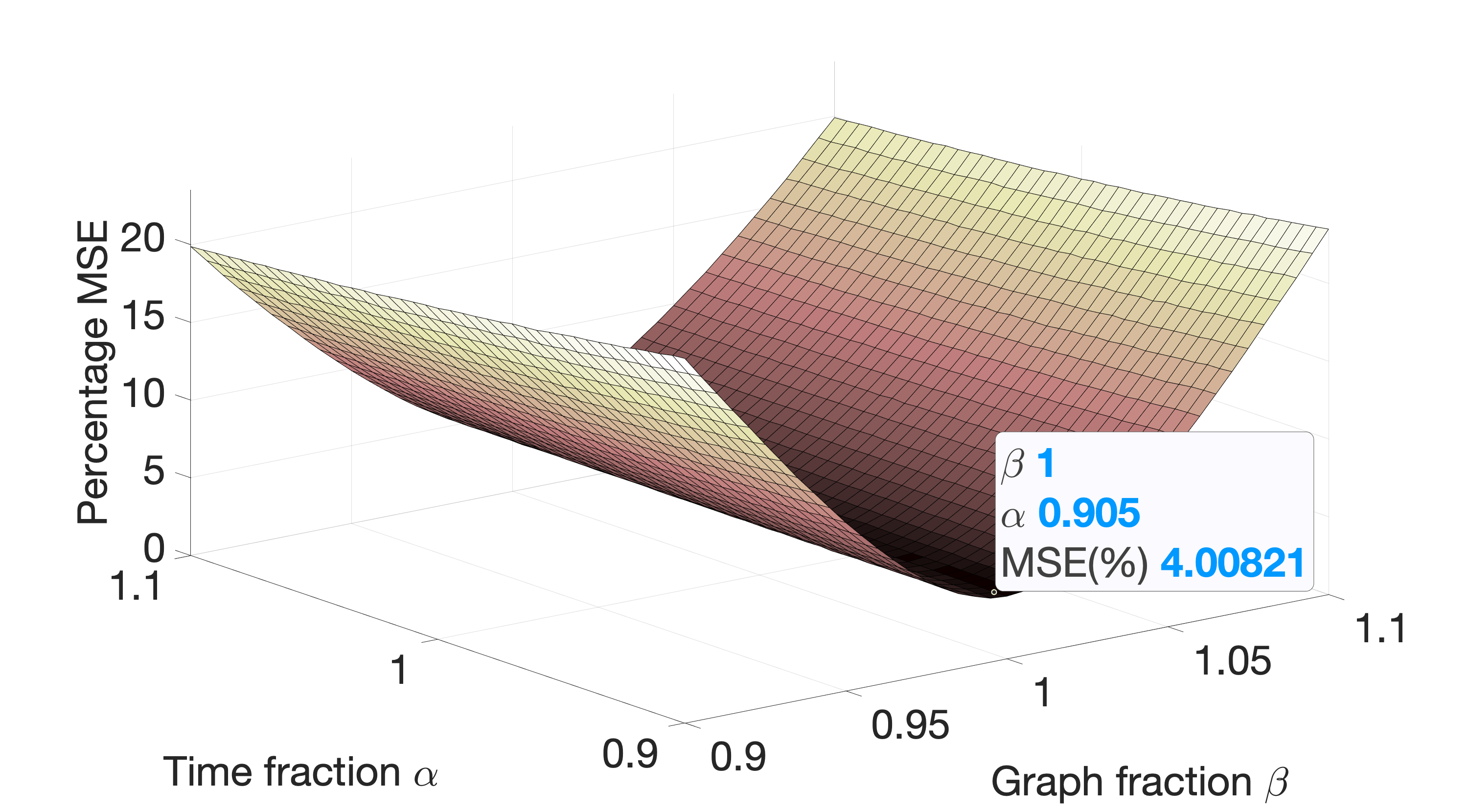}
	\end{subfigure}
	\caption{The results of the regularization-based denoising applied to the Molene dataset. (Left) varying regularization parameters are considered while fractional orders are fixed to $\alpha = 0.905$ and $\beta =1$. (Right) fractional orders are considered with fixed regularization parameters $\tau_g = 3.8, \tau_t = 4$.}~\label{weather_molene}
\end{figure}

\begin{figure}[ht]
	\centering
	\begin{subfigure}[t]{0.49\linewidth}
		\centering
		\includegraphics[width=\linewidth, height=0.25\textheight, keepaspectratio]{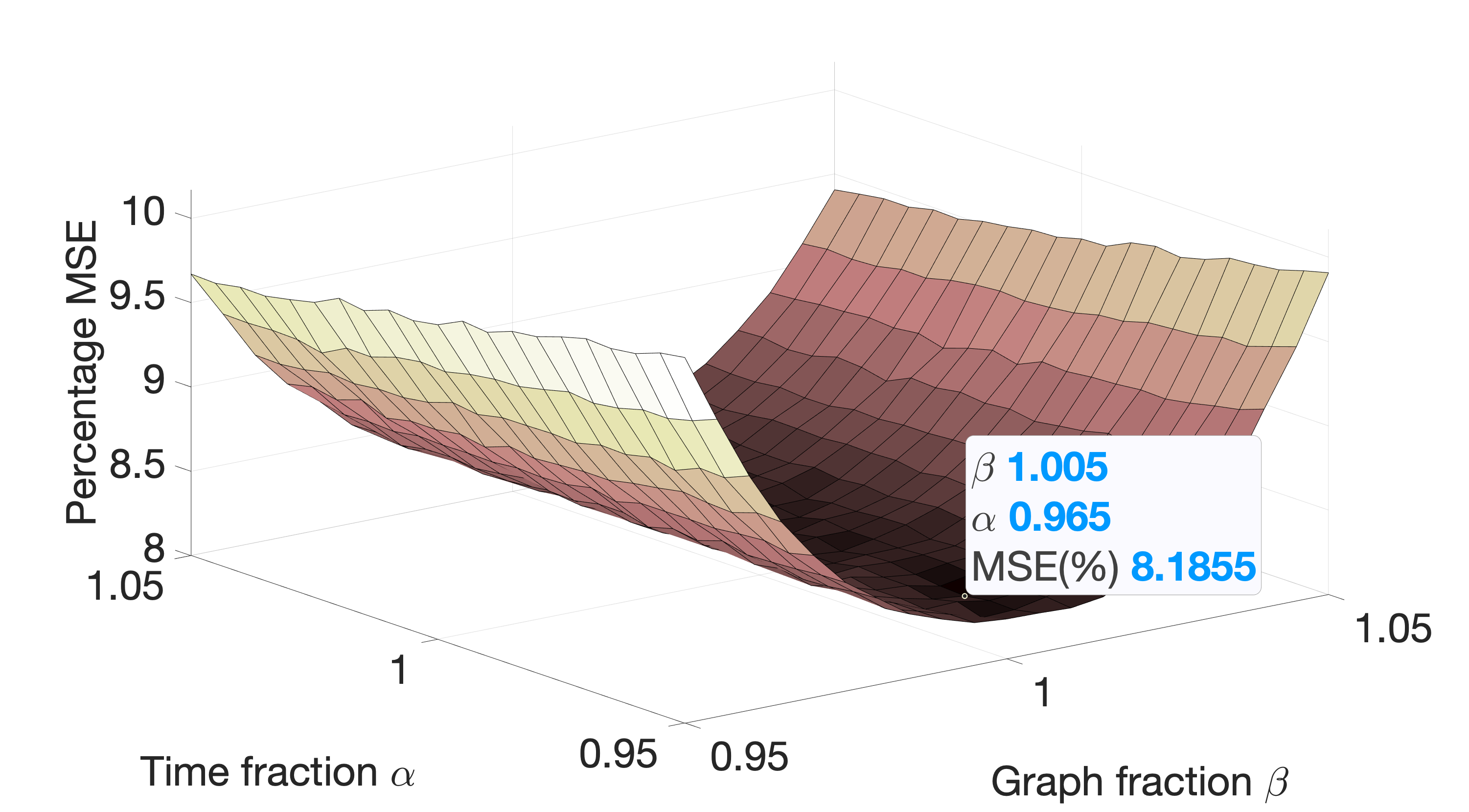}
	\end{subfigure}
	\begin{subfigure}[t]{0.49\linewidth}
		\centering
		\includegraphics[width=\linewidth, height=0.25\textheight, keepaspectratio]{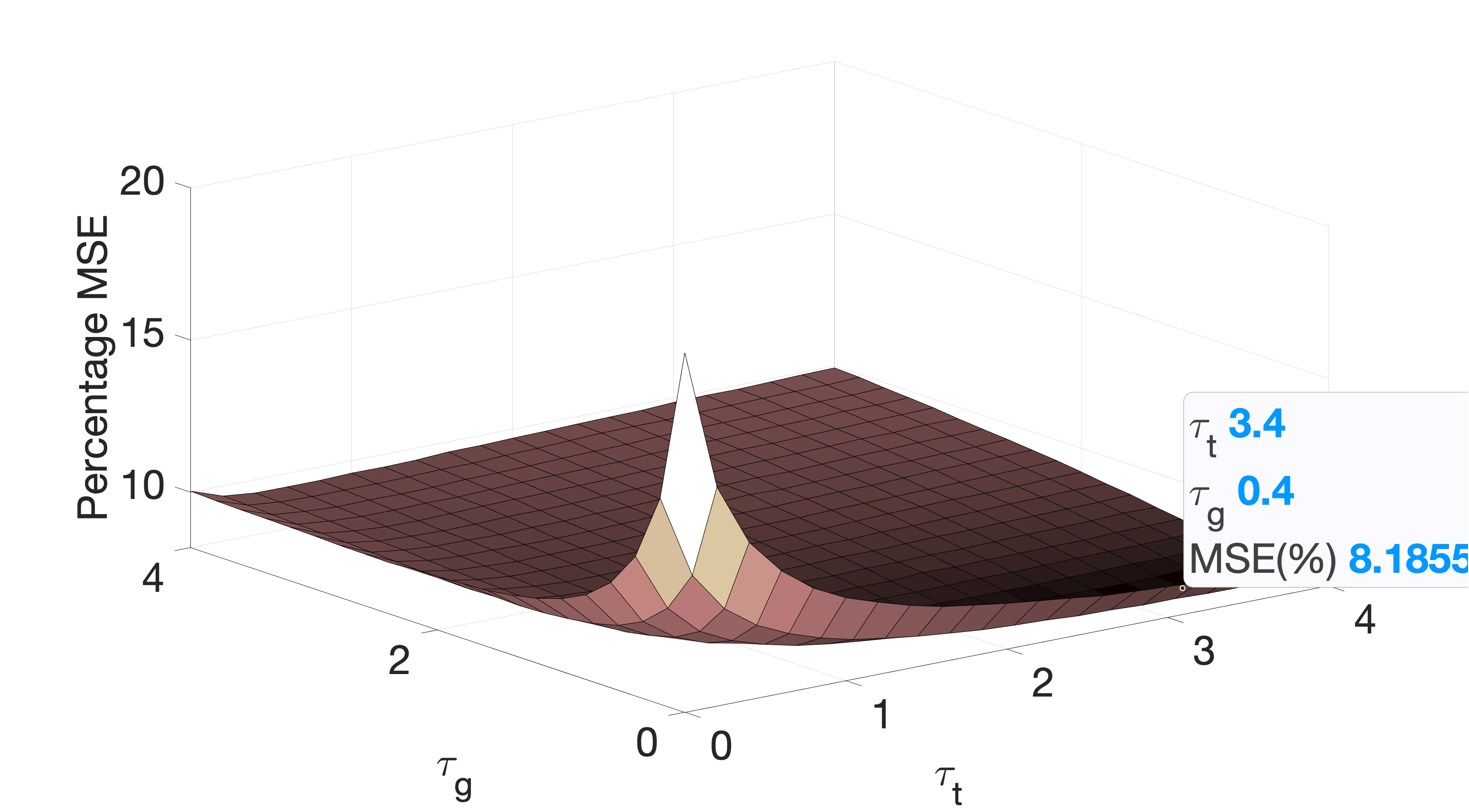}
	\end{subfigure}
	\\
	\begin{subfigure}[t]{0.49\linewidth}
		\centering
		\includegraphics[width=\linewidth, height=0.25\textheight, keepaspectratio]{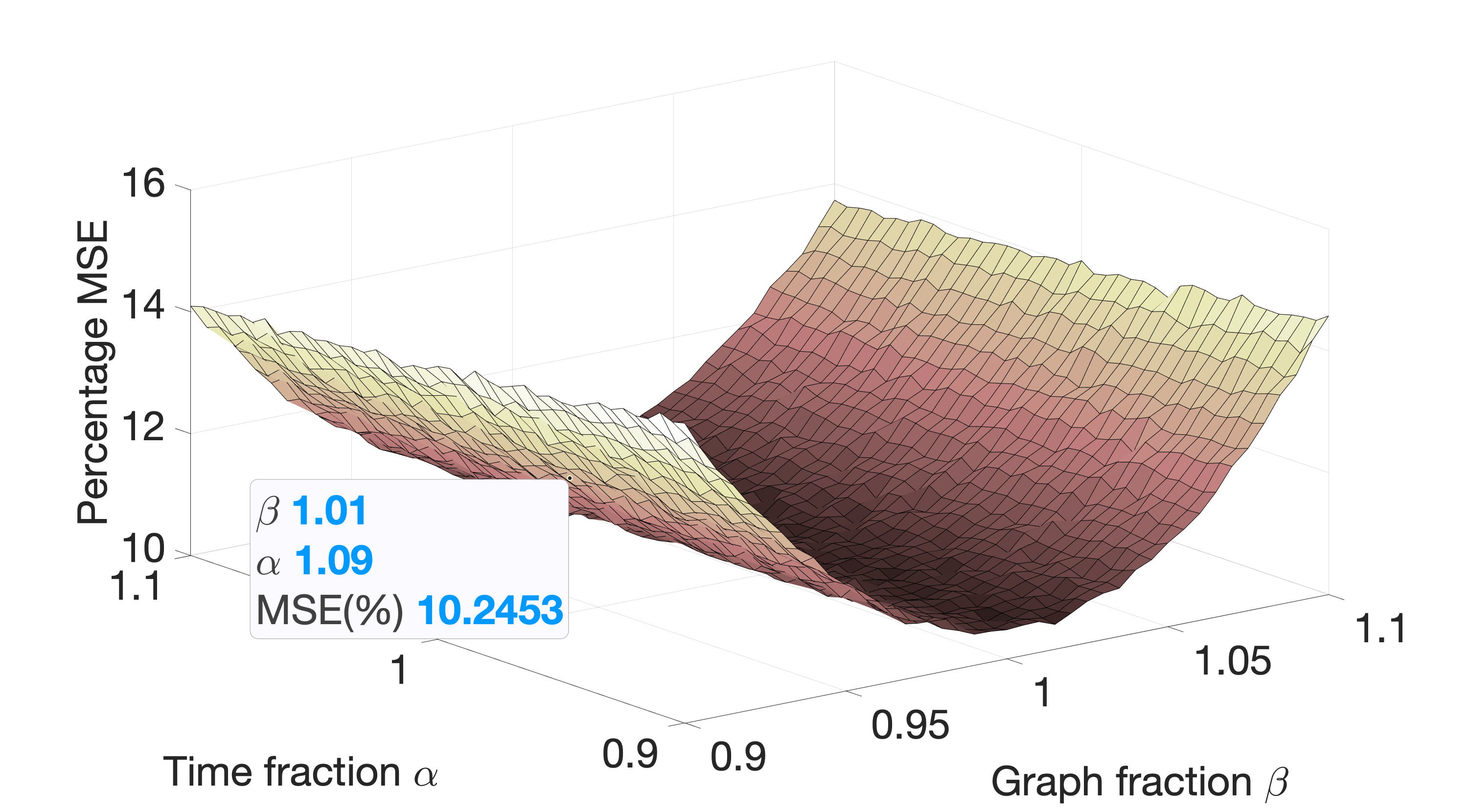}
	\end{subfigure}
	\begin{subfigure}[t]{0.49\linewidth}
		\includegraphics[width=\linewidth, height=0.25\textheight, keepaspectratio]{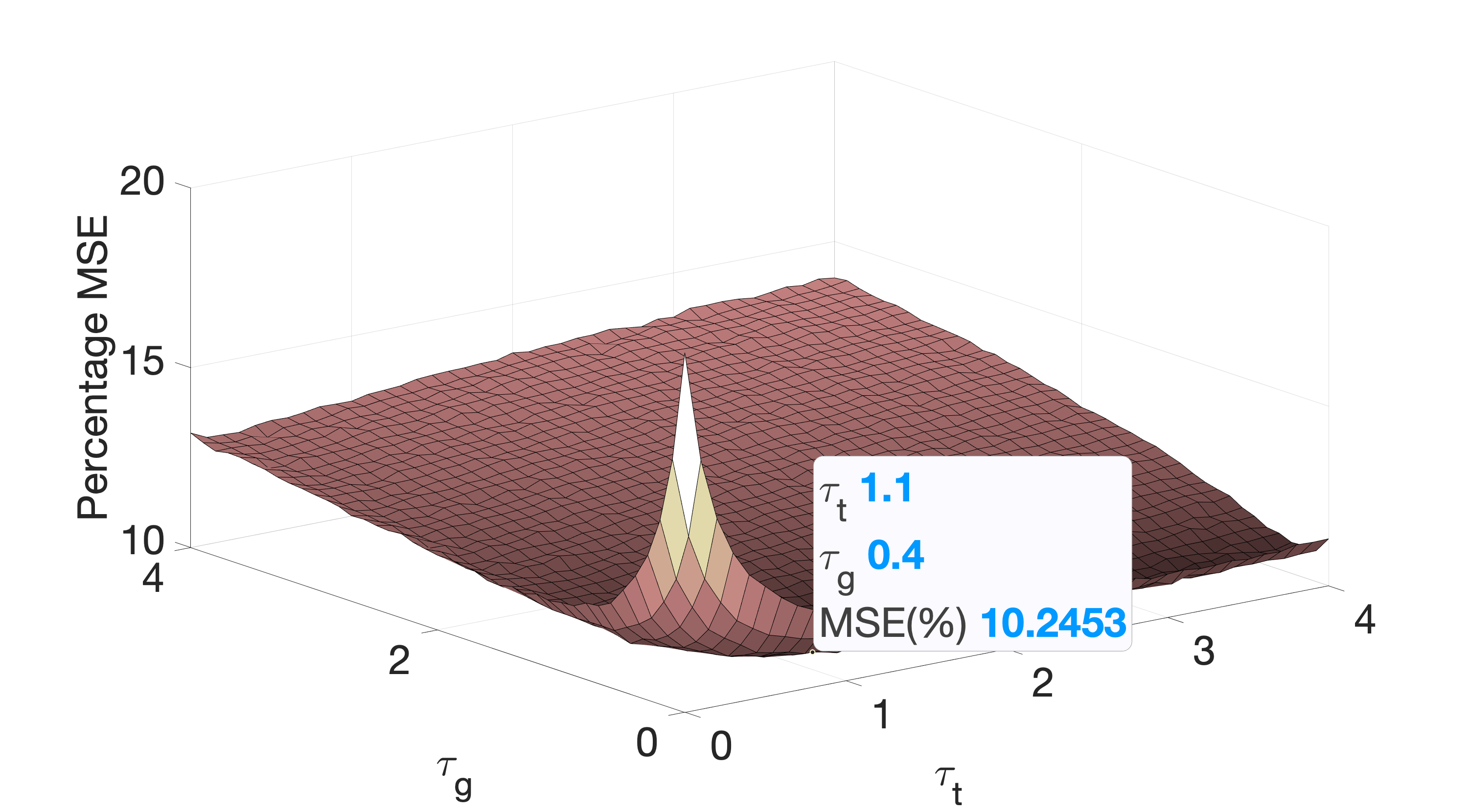}
	\end{subfigure}
	\caption{\small In the NOAA dataset, both yearly (top row) and monthly (bottom row) settings are considered. Performances of denoising with respect to (Upper left) varying $\alpha,\beta$ with fixed $\tau_t = 3.4,\tau_g = 0.4$. (Upper right) varying $\tau_t,\tau_g$ with fixed $\alpha = 0.965, \beta = 1.005$. (Lower left) varying $\alpha,\beta$ with fixed $\tau_t = 1.1, \tau_g = 0.4$ and (Lower right) varying $\tau_t,\tau_g$ with fixed $\alpha = 1.09, \beta = 1.01$.}~\label{weather_station}
\end{figure}

The results for the Molene dataset are presented in Fig.~\ref{weather_molene}. Fig.~\ref{weather_station} (top row) gives the results for the one-year cycle of the NOAA dataset. We also explored a larger span of fractional orders and regularization parameters for one month (Jan. 2020), and the results are presented in Fig.~\ref{weather_station} (bottom row). It can be inferred from these results that for one-year cycle data, the minimum percentage mean squared error (MSE) is obtained at order pair $(0.965,1.005)$  with regularization parameters $\tau_g = 0.4$ and $\tau_t = 3.4$. The minimum percentage MSE is obtained for monthly data at order $(1.09,1.01)$ with $\tau_g = 0.4$ and $\tau_t = 1.1$. For the Molene dataset, the minimum is obtained at order $(0.905,1.0)$ with $\tau_t = 4$ and $\tau_g = 3.8$. Hence, the results suggest that JFRT for filtering in the fractional orders for both domains improves performance compared to ordinary JFT\@.

\subsection{Clustering Experiments}\label{sec:exp:clustering}
\begin{figure}[ht]
	\centering
	\begin{subfigure}[t]{0.32\linewidth}
		\centering
		\includegraphics[width=\linewidth]{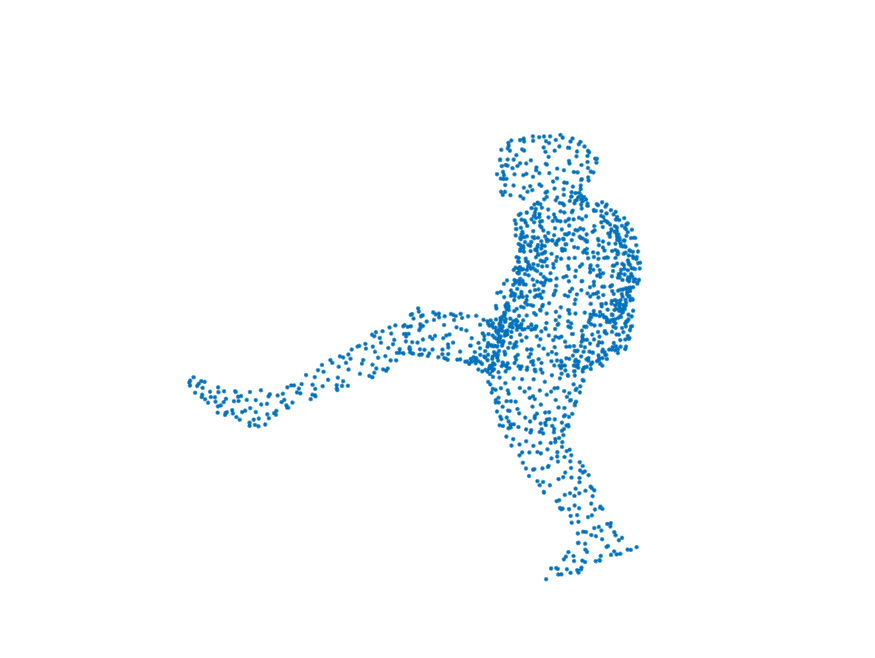}
	\end{subfigure}
	\begin{subfigure}[t]{0.32\linewidth}
		\centering
		\includegraphics[width=\linewidth]{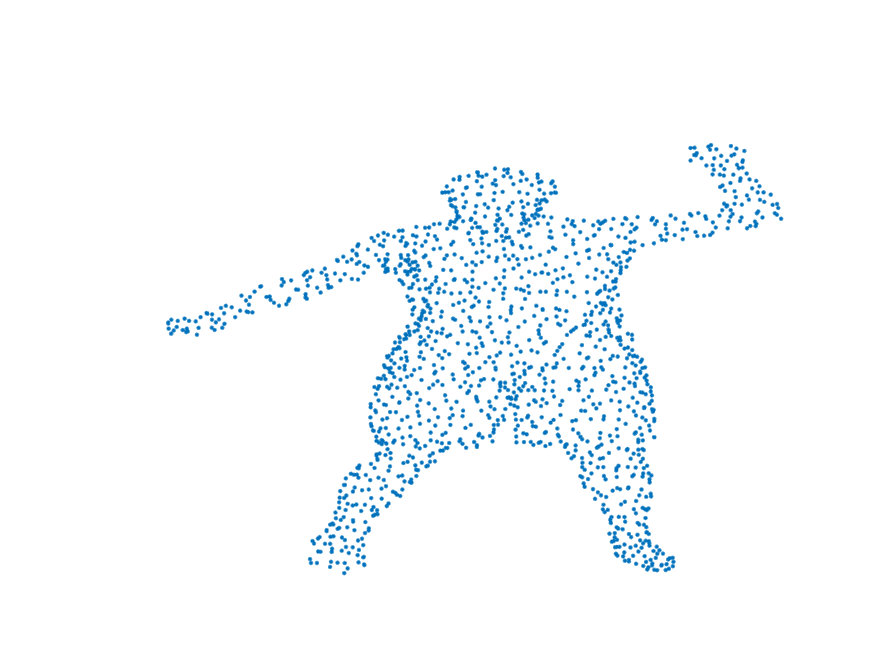}
	\end{subfigure}
	\begin{subfigure}[t]{0.32\linewidth}
		\centering
		\includegraphics[width=\linewidth]{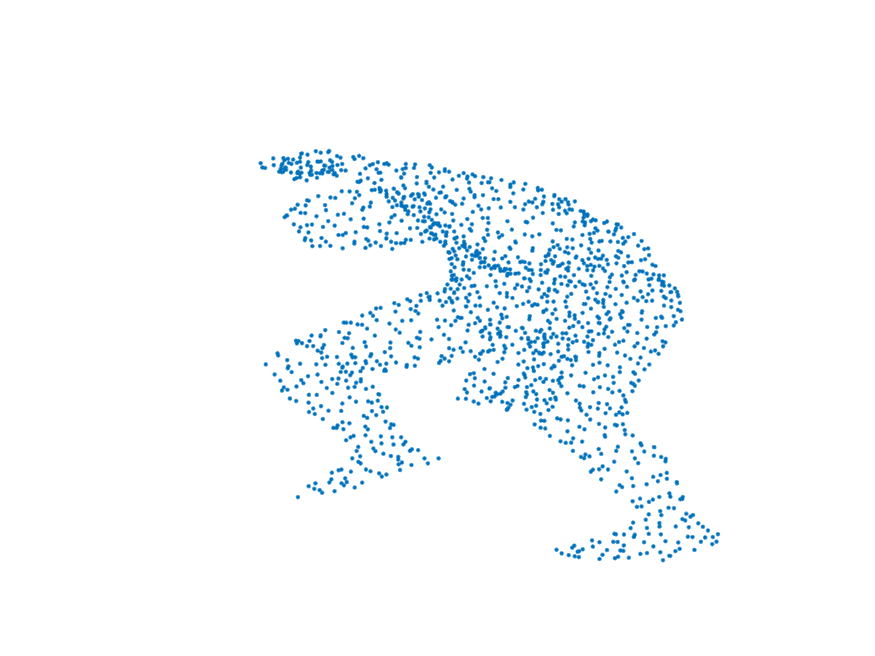}
	\end{subfigure}
	\caption{Sample frames from the actions of the Dancer mesh. From left-to-right, \textit{stretching legs}, \textit{moving arms} and \textit{bending body}.}\label{mesh_points}
\end{figure}

For this experiment, we consider the motions of \textit{the Dancer mesh}, which consists of 1,502 coordinates in 3D space along 573-time samples~\cite{grassi18timevertex}. These motions are \textit{moving arms}, \textit{stretching legs}, and \textit{bending body} motions. 2D plots of some actions can be seen in Fig.~\ref{mesh_points}. We follow the experimental procedure of~\cite{grassi18timevertex} to demonstrate the performance and utility of JFRT\@. The mesh is corrupted with additive sparse noise density \(0.1\), meaning that 10\% of mesh points are corrupted. The noise is Gaussian with a SNR of \(-10\,\DB \) and \(-20\,\DB \) as in~\cite{grassi18timevertex}. We use windows of size \(50\) with $60\%$ overlap to obtain 27-time sequences and $k$-NN graph structure to capture the geometry.

\begin{figure}[ht]
	\centering{}
	\begin{subfigure}{0.49\linewidth}
		\centering{}
		\includegraphics[width=\linewidth]{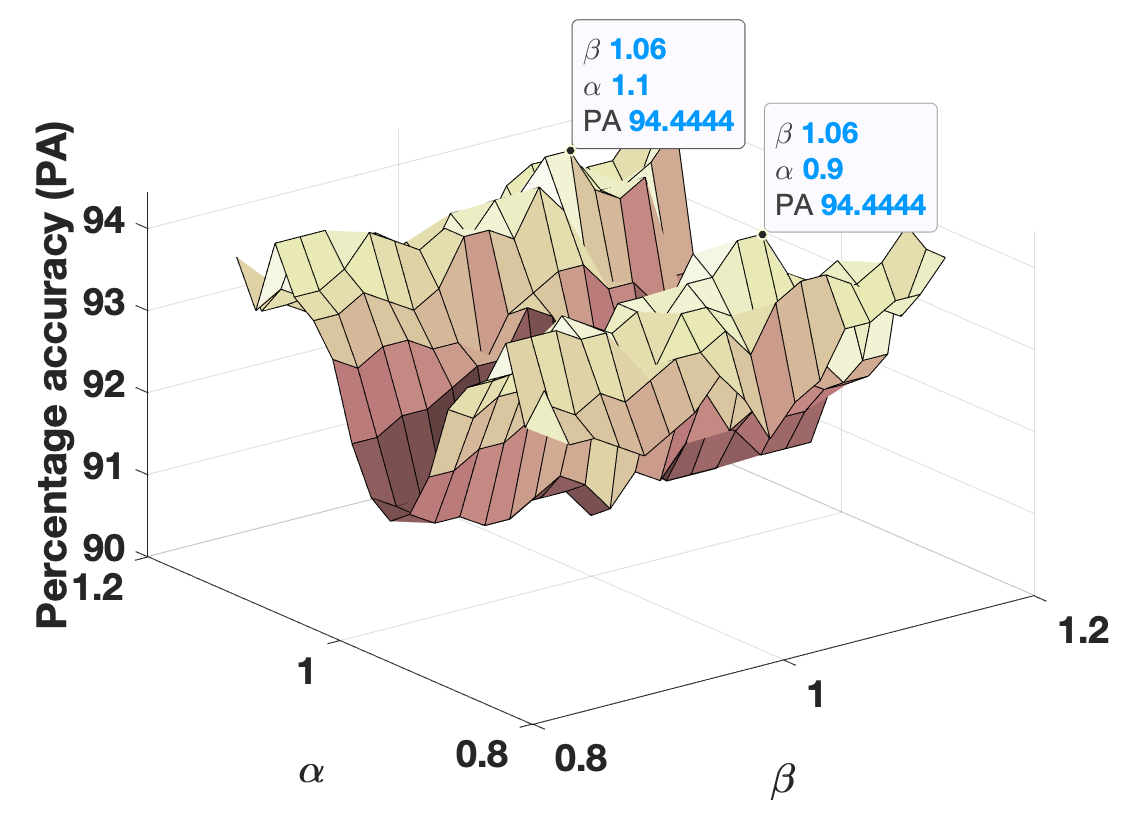}
	\end{subfigure}
	\begin{subfigure}{0.49\linewidth}
		\centering{}
		\includegraphics[width=\linewidth]{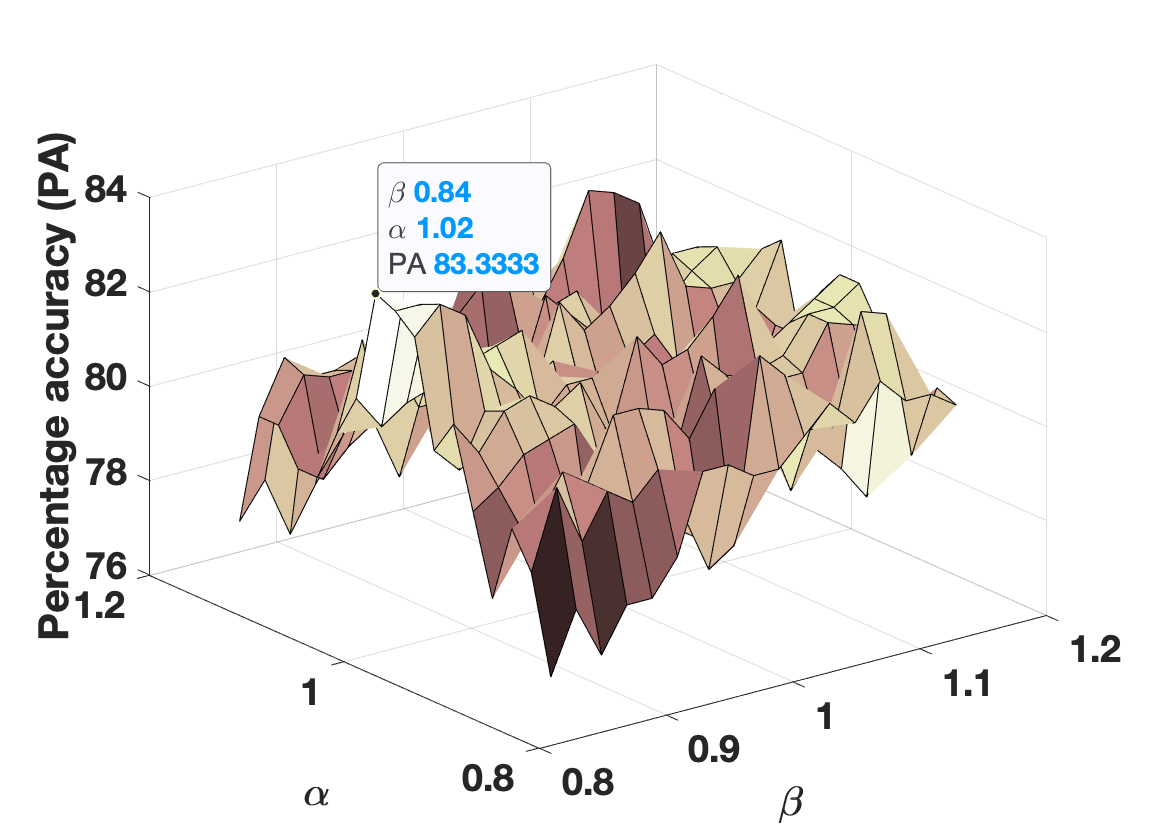}
	\end{subfigure}
	\caption{Performance of clustering accuracy for varying JFRT with (left) \(-10\,\DB \) SNR and (right) \(-20\,\DB \) SNR.}~\label{accclus}
\end{figure}

We calculate the JFRT of the obtained sequences in each of the coordinate dimensions, concatenate the resulting fractional order joint time-vertex signals, and finally cluster the resulting representation to get the classifications by using the \(k\)-means clustering algorithm (repeated 20 times). We present the results for the average accuracy of different JFRT orders for \(-10\) and \(-20\,\DB \) SNR in Fig.~\ref{accclus}. We extend the search range for the $-20\,\DB$ SNR experiment in Fig.~\ref{accclus} to $\alpha,\beta\in [0, 4]$ with step sizes of $0.02$. With the broader range, we find a new setting of $(\alpha, \beta)=(0.74, 1.44)$, which increased the accuracy from the previous range's $83.33\%$ to $85.19\%$ as depicted below in Fig~\ref{fig:large-range}. We highlight that our main aim is to show we can obtain a better result at $(\alpha, \beta)\neq(1, 1)$. We would like to remind that the optimal $(\alpha, \beta)$ values depend on the given data and its inherent characteristics, and one can find a better solution with a much broader range and with more resolution. We achieve the highest accuracy levels at fractional orders for both cases, indicating performance improvements over ordinary JFT\@.
\begin{figure}[ht]
	\centering
	\includegraphics[width=\linewidth]{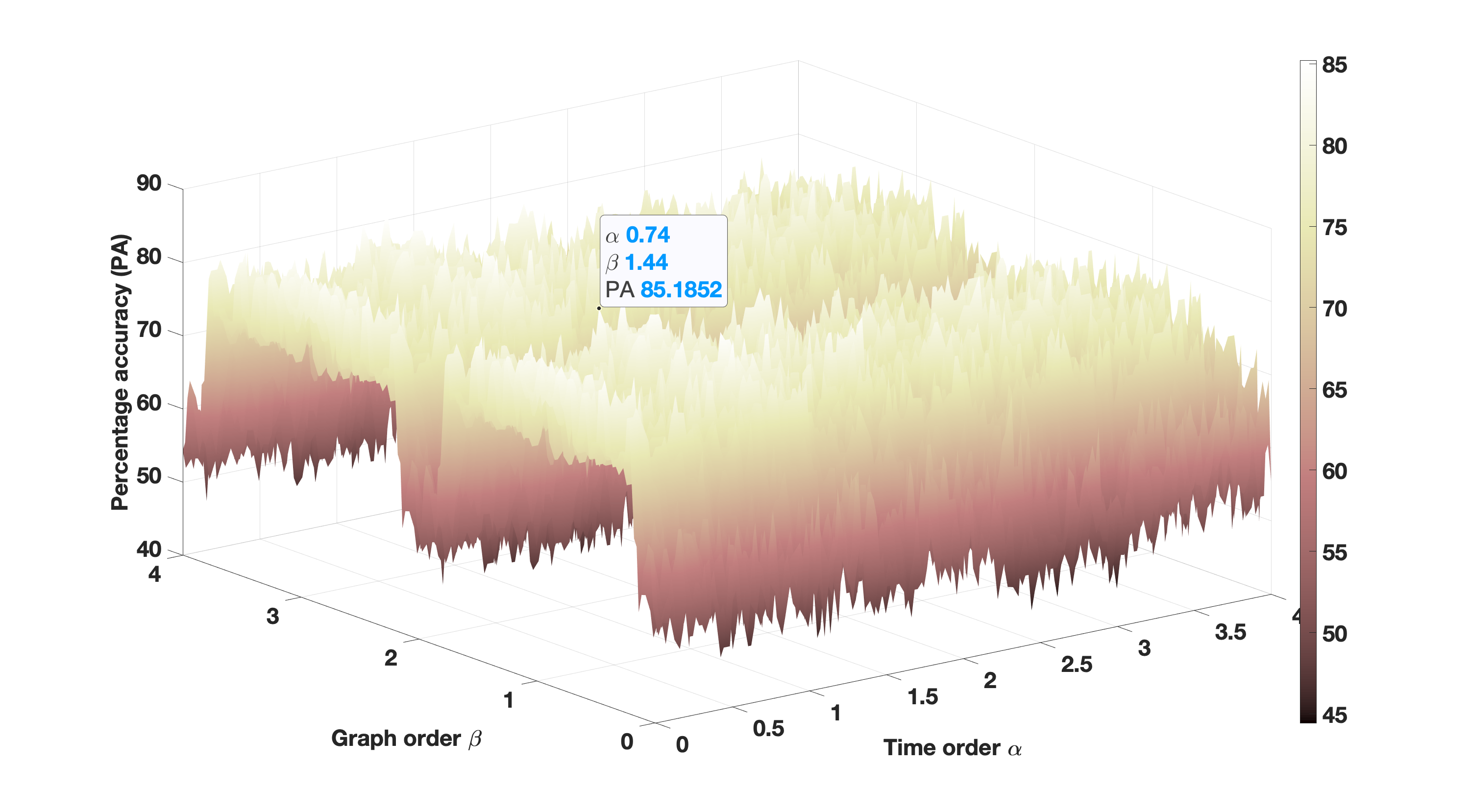}
	\caption{Performance of clustering accuracy for varying JFRT with \(-20\,\DB \) SNR on the extended range of $\alpha,\beta\in [0, 4]$ with step sizes of $0.02$.}\label{fig:large-range}
\end{figure}
\begin{figure}[ht]
	\centering
	\begin{subfigure}{\linewidth}
		\centering\includegraphics[width=\linewidth]{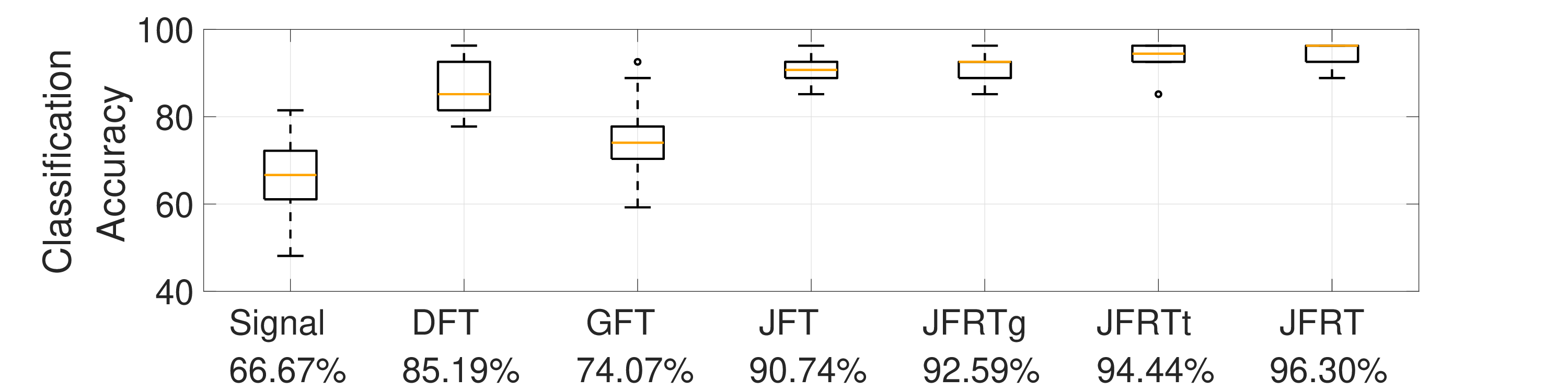}
	\end{subfigure}
	\begin{subfigure}{\linewidth}
		\centering\includegraphics[width=\linewidth]{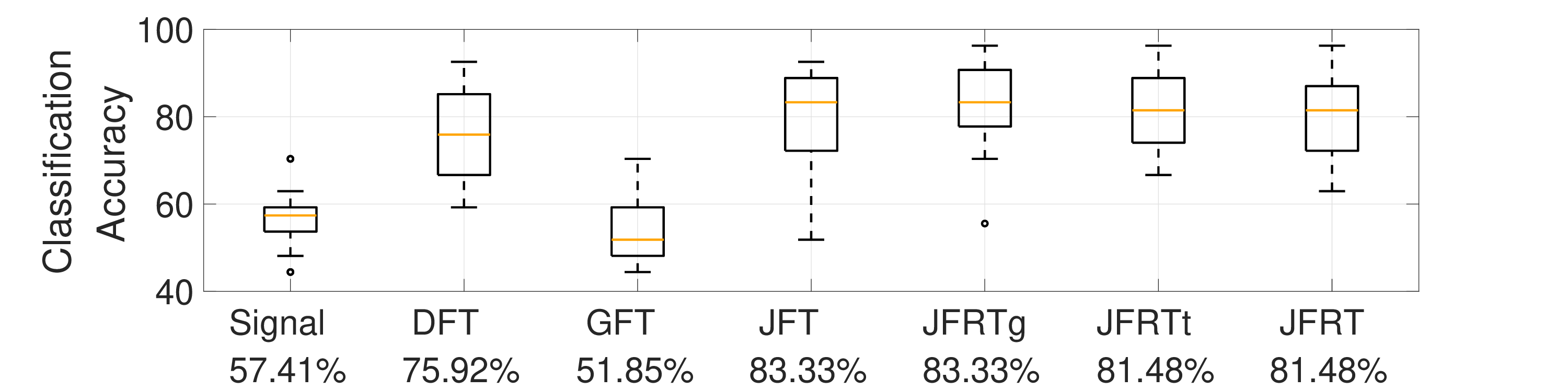}
	\end{subfigure}
	\caption{Box plots for the clustering performance of 20 samples for the original signal, DFT, GFT, JFT, JFRTg, JFRTt, JFRT, with (top) \(-10\,\DB \) SNR and (bottom) \(-20\,\DB \) SNR\@. The horizontal lines in the box plots denote the means of distribution.}\label{box}
\end{figure}

To make further comparisons, we also considered the cases of the ordinary JFT, the best obtained average accuracy of the (time-ordinary)\slash(graph-fractional) Fourier transform (JFRTg), the best obtained average accuracy of the (graph-ordinary)\slash(time-fractional) Fourier transform (JFRTt), the best obtained average accuracy for arbitrary JFRT and the received mesh (signal) without any transformation. JFRTg and JFRTt are the cases where we only let the graph order be fractional and the time order be fractional, respectively. We also repeat this experiment 20 times, and the results for \(-10\,\DB \) SNR and \(-20\,\DB \) SNR in box plots are provided in Fig.~\ref{box}. For both noise levels, the accuracy distribution is more confined to higher percentages at fractional orders, and we achieve the best mean accuracy at fractional orders. Specifically, JFRT provides $3.7\%$ and $2.6\%$ improvements of mean classification accuracy over JFT for \(-10\,\DB \) and \(-20\,\DB \) SNR, respectively. These results show that JFRT provides better clustering performance as the mesh points of similar motions are densely populated.

\section{Conclusion}\label{conclusion}
We proposed the JFRT, a fractional transformation for processing joint time-vertex signals as a generalization to ordinary JFT and two-dimensional DFRT\@. With JFRT, it is possible to jointly transform time-varying graph signals into domains between vertex and graph-spectral domains from the graph perspective and between time and frequency domains from the ordinary time-series perspective. Thus, JFRT could be seen as a transformation for two-dimensional joint time-vertex signals. We showed that the proposed JFRT is index additive in orders, reversible, and commutative. We also showed that the JFRT is unitary if the underlying GFT is unitary. JFRT reduces to the ordinary JFT when the order is $(1,1)$, reduces to identity when the order is $(0,0)$, and, for directed circular graphs, reduces to the ordinary two-dimensional DFRT for the order $(1,1)$. This makes JFT, 2D DFRT, and 2D-DFT special cases of the JFRT\@. We proposed fractional filtering for the joint time-vertex signals based on JFRT and showed that the optimal JFRT parameters differ from $(1,1)$, corresponding to the ordinary JFT\@. We also constructed Tikhonov regularization-based denoising in the proposed JFRT domains using the joint fractional Laplacian to regularize a received signal separately in both fractional time and fractional graph domains. We derived the associated optimal filter coefficients to be used in denoising.

The extra flexibility that the JFRT provides through its two parameters without imposing additional computational costs can open up several performance increases over the non-parametric JFT in joint time-vertex signal processing. We provided numerical examples of denoising and clustering tasks such that JFRT allows us to improve performance. As JFRT offers a new and flexible framework to handle joint-time vertex data, we expect it will be instrumental for several signal processing applications and open new theoretical research directions in the joint time-vertex GSP\@. On the other side of the coin, we also extended the literature on FRT and contributed to the generalizations from the classical FRT analysis to the GSP domain by introducing JFRT\@.

\section*{Acknowledgments}
This work was supported by T\"urk Telekom\"unikasyon A.S. through the 5G and Beyond Joint Graduate Support Program coordinated by the Information and Communication Technologies Authority and also with TUBITAK 1001 Grant (124E179). The work of Aykut Ko\c c was supported by the BAGEP Award of the Science Academy.

\bibliographystyle{elsarticle-num}
\bibliography{bibliography}

\appendix
\section{Time-Varying Graph Signal Median Filters}\label{sec:app:median}
The median filtering approach is relatively straightforward. With the \(\text{Median}_1\) approach, an entry of a joint time-vertex signal is replaced with the median of the entries located at the neighbors of the given vertex at the current time, as well as the entries of the same vertex at previous, current and next time instances~\cite{tay21timevaryingdenoising}. With the \(\text{Median}_{2}\) approach, in addition to the entries used in \(\text{Median}_1\) approach, the entries at the neighbors of the current vertex at the previous and next time instances are also used~\cite{tay21timevaryingdenoising}. We implement this approach based on the provided definitions.

\section{Autoregressive Moving Average Graph Filters}\label{sec:app:arma}
The ARMA recursion for underlying static graph but with time-varying graph signal is provided as follows~\cite{isufi17armafilter}:
\begin{align}
	\boldsymbol{y}_{t+1} = \psi\boldsymbol{L} \boldsymbol{y}_t + \varphi \boldsymbol{x}_t,\quad
	\boldsymbol{z}_{t+1} = \boldsymbol{y}_{t+1} + c \boldsymbol{x}_t,
\end{align}
where \(\boldsymbol{x}_t\), \(\boldsymbol{z}_t\), \(\boldsymbol{y}_t\) are the graph signal to be filtered, filtered graph signal and intermediate graph signal at time \(t\), respectively, \(\boldsymbol{L}\) is the graph Laplacian, \(\psi,\varphi,c\in\mathbb{C}\) are the ARMA coefficients to be determined. With \(K\)-parallel ARMA graph filter, the given recursion results in the following joint graph and temporal frequency transfer function~\cite{isufi17armafilter}:
\begin{align}
	H(z,\lambda) & = \sum_{k=1}^{K}\frac{\varphi^{(k)} z^{-1}}{1 - \psi^{(k)}\lambda z^{-1}} + c z^{-1}.
\end{align}
The coefficients \(\psi^{(k)}\) and \(\varphi^{(k)}\) are defined based on the poles (\(p_k\)) and residues (\(r_k\)) of the transfer function,
\begin{equation}
	\psi^{(k)} = \frac{1}{p_k},\quad\varphi^{(k)} = -\frac{r_k}{p_k},
\end{equation}
described through the partial fraction decomposition of the transfer function given in terms of \(a_k\) and \(b_k\)  coefficients as
\begin{equation}
	\frac{\sum_{k=1}^{K}b_k s^{K-k}}{\sum_{k=1}^{K}a_k s^{K-k}} = c(s) + \sum_{k=1}^{K}\frac{r_k}{s-p_k}.
\end{equation}

Since the ARMA graph filters are generated through \(a_k\) and \(b_k\) coefficients, instead of \(r_k\) and \(p_k\) values, these coefficients are provided as opposed to~\cite{isufi17armafilter}. Since we can only access the source code of the static graph signal ARMA filtering method, we implement the time-varying version based on the definitions provided in~\cite{isufi17armafilter}, and the partial source code\footnote{\href{https://andreasloukas.blog/code/}{https://andreasloukas.blog/code}}. We obtain the ARMA graph filter coefficients by using the provided source code and experiment with several \textit{parallel \(\text{ARMA}_K\) filters} since they are the best ARMA graph filtering approach for time-varying graph signals~\cite{isufi17armafilter}.

\section[Untrained GNN]{Untrained GNN}\label{sec:untrained}
The untrainded GNN framework~\cite{rey22untrainedgnn} is defined for a graph $\mathcal{G}$, any GNN architecture parametrized with set of learnable weights $\mathbf{W}$ and input $\mathbf{X}_0$, $\text{GNN}_{\mathbf{W}}(\mathbf{X}_0\mid\mathcal{G})$, joint time-vertex signal and noise $\mathbf{X},\mathbf{N}$ and noisy signal $\mathbf{Y} = \mathbf{X} + \mathbf{N}$. The training approach is to feed the network with i.i.d.\ zero-mean Gaussian noise such that ${(\mathbf{N}_{\text{input}})}_{i,j}\sim\mathcal{N}(0, \sigma^2)$ find the optimal set of weights $\widehat{\mathbf{W}}$ that minimizes the loss function provided in~\cref{eq:untrained_loss}. Then, with the trained weights, get a denoised prediction of the original signal as in~\cref{eq:untrained_prediction}.
\begin{align}
	\label{eq:untrained_loss}
	\mathcal{L}(\mathbf{Y}, \mathbf{W}) & = \frac{1}{2}\NORM{\mathbf{Y} - \text{GNN}_{\mathbf{W}}(\mathbf{N}_{\text{input}}\mid\mathcal{G})}{F}^2 \\
	\label{eq:untrained_prediction}
	\widehat{\mathbf{X}}                & = \text{GNN}_{\widehat{\mathbf{W}}}(\mathbf{N}_{\text{input}}\mid\mathcal{G})
\end{align}
The network is called ``untrained'' because the parameters of the network are optimized using only the signal observation that is required to be denoised, avoiding the dependency on a training set with several observed graph signals.

\section[TimeGNN]{TimeGNN}\label{sec:timegnn}
We adjusted the TimeGNN architecture that is proposed in~\cite{castro23timegnn}. Even though it is proposed as a ``recovery'' and not a ``denoising'' approach, its architecture is applicable to our problem with the untrained GNN framework~\cite{rey22untrainedgnn}. The TimeGNN architecture is based on the Chebyshev spectral graph convolutional operator with the following layer update rule [2]: For the $\ell^{\text{th}}$-layer input $\mathbf{H}^{(\ell)}$, the shifted matrices $\boldsymbol{\mathcal{Z}}_{k}^{(\ell)}$ are calculated recursively as $\boldsymbol{\mathcal{Z}}_{k}^{(\ell)} = 2\widehat{\mathbf{L}}\boldsymbol{\mathcal{Z}}_{k - 1}^{(\ell)} - \boldsymbol{\mathcal{Z}}_{k - 2}^{(\ell)}$, where $\boldsymbol{\mathcal{Z}}_{1}^{(\ell)} = \mathbf{H}^{(\ell)}$ and $ \boldsymbol{\mathcal{Z}}_{2}^{(\ell)} = \widehat{\mathbf{L}}\mathbf{H}^{(\ell)}$. In this context, $\widehat{\mathbf{L}} = \frac{2}{\lambda_{\max}}\mathbf{L} - \mathbf{I}$ is the normalized Laplacian with eigenvalues reside in $[-1, 1]$. With these definitions, the next layer output is calculated as follows:
\begin{equation}
	\mathbf{H}^{(\ell + 1)} = \sum_{\rho=1}^{\alpha}\mu_{\rho}^{(\ell)}\sum_{k=1}^{\rho}\boldsymbol{\mathcal{Z}}_{k}^{(\ell)}\mathbf{W}_{k,\rho}^{(\ell)},
\end{equation}
where $\mu_{\rho}^{(\ell)}$ are the learnable weights for the linear $\ell^{\text{th}}$-layer's linear combination and $\alpha$ is a hyperparameter that determines the number of Chebyshev branches.

\section{Experiments}\label{sec:app:experiments}
\begin{table}[ht]
	\centering
	\caption{The best-performing JFRT settings for the denoising experiment on synthetic data, where \(d\) is the delay multiplier, \(\sigma \) is Gaussian noise standard deviation, \((e_n,e_e)\) are the noise and estimation errors, respectively, \(\alpha \) is the FRT fraction, \(\beta \) is the GFRT fraction, and \(c\) is the number of trailing zeros in the low-pass graph filter.}\label{tab:denoise:synthetic:best}
	\resizebox{\textwidth}{!}{
		\begin{tabular}{|c||llllllllr||llllllllr||lllllllll||}\hline{}
			\multirow{3}{*}{\(d\)} & \multicolumn{9}{c||}{\(\sigma = 0.10\)} & \multicolumn{9}{c||}{\(\sigma = 0.15\)} & \multicolumn{9}{c||}{\(\sigma = 0.20\)}                                                                                                                                                                                                                                                                                                                                                                                                                                                                                                                                                                                                                                                                                                                                                                                                  \\ \cline{2-28}
			                       & \multicolumn{1}{c|}{}                   & \multicolumn{4}{c|}{Adjacency}          & \multicolumn{4}{c||}{Laplacian}         & \multicolumn{1}{c|}{}          & \multicolumn{4}{c|}{Adjacency} & \multicolumn{4}{c||}{Laplacian} & \multicolumn{1}{c|}{}           & \multicolumn{4}{c|}{Adjacency} & \multicolumn{4}{c||}{Laplacian}                                                                                                                                                                                                                                                                                                                                                                                                                                                                                                                                                                                         \\ \cline{2-28}
			                       & \multicolumn{1}{c|}{\(e_n\)}            & \multicolumn{1}{c}{\(e_e\)}             & \multicolumn{1}{c}{\(\alpha \)}         & \multicolumn{1}{c}{\(\beta \)} & \multicolumn{1}{c|}{\(c\)}     & \multicolumn{1}{c}{\(e_e\)}     & \multicolumn{1}{c}{\(\alpha \)} & \multicolumn{1}{c}{\(\beta \)} & \multicolumn{1}{c||}{\(c\)}     & \multicolumn{1}{c|}{\(e_n\)} & \multicolumn{1}{c}{\(e_e\)} & \multicolumn{1}{c}{\(\alpha \)} & \multicolumn{1}{c}{\(\beta \)} & \multicolumn{1}{c|}{\(c\)} & \multicolumn{1}{c}{\(e_e\)} & \multicolumn{1}{c}{\(\alpha \)} & \multicolumn{1}{c}{\(\beta \)} & \multicolumn{1}{c||}{\(c\)} & \multicolumn{1}{c|}{\(e_n\)} & \multicolumn{1}{c}{\(e_e\)} & \multicolumn{1}{c}{\(\alpha \)} & \multicolumn{1}{c}{\(\beta \)} & \multicolumn{1}{c|}{\(c\)} & \multicolumn{1}{c}{\(e_e\)} & \multicolumn{1}{c}{\(\alpha \)} & \multicolumn{1}{c}{\(\beta \)} & \multicolumn{1}{c||}{\(c\)} \\ \hline
			\(25\)                 & \multicolumn{1}{l|}{14.13}              & 11.37                                   & 1.33                                    & 1.00                           & \multicolumn{1}{r|}{35}        & 11.60                           & 1.33                            & 1.00                           & 26                              & \multicolumn{1}{l|}{21.19}   & 15.48                       & 1.31                            & 1.00                           & \multicolumn{1}{r|}{35}    & 16.02                       & 1.32                            & 1.01                           & 46                          & \multicolumn{1}{l|}{28.25}   & 18.27                       & 1.34                            & 1.00                           & \multicolumn{1}{l|}{49}    & 18.31                       & 1.34                            & 1.01                           & 49                          \\
			\(30\)                 & \multicolumn{1}{l|}{14.24}              & 12.54                                   & 0.69                                    & 1.01                           & \multicolumn{1}{r|}{12}        & 12.20                           & 1.32                            & 1.00                           & 18                              & \multicolumn{1}{l|}{21.36}   & 16.42                       & 1.34                            & 1.01                           & \multicolumn{1}{r|}{35}    & 16.95                       & 1.23                            & 1.00                           & 26                          & \multicolumn{1}{l|}{28.48}   & 20.30                       & 1.34                            & 1.01                           & \multicolumn{1}{l|}{35}    & 20.55                       & 1.34                            & 1.01                           & 46                          \\
			\(35\)                 & \multicolumn{1}{l|}{14.30}              & 12.98                                   & 0.69                                    & 1.02                           & \multicolumn{1}{r|}{12}        & 12.75                           & 0.68                            & 1.00                           & 9                               & \multicolumn{1}{l|}{21.45}   & 17.99                       & 1.35                            & 1.01                           & \multicolumn{1}{r|}{31}    & 18.08                       & 1.24                            & 0.99                           & 19                          & \multicolumn{1}{l|}{28.59}   & 21.57                       & 1.34                            & 1.01                           & \multicolumn{1}{l|}{35}    & 22.15                       & 1.34                            & 1.00                           & 26                          \\
			\(40\)                 & \multicolumn{1}{l|}{14.35}              & 13.23                                   & 0.68                                    & 1.01                           & \multicolumn{1}{r|}{1}         & 13.05                           & 0.68                            & 1.02                           & 4                               & \multicolumn{1}{l|}{21.53}   & 19.21                       & 1.36                            & 1.02                           & \multicolumn{1}{r|}{11}    & 18.63                       & 1.35                            & 0.99                           & 18                          & \multicolumn{1}{l|}{28.69}   & 23.15                       & 1.34                            & 1.01                           & \multicolumn{1}{l|}{31}    & 23.12                       & 1.34                            & 1.00                           & 26                          \\
			\(45\)                 & \multicolumn{1}{l|}{14.33}              & 13.22                                   & 0.68                                    & 1.02                           & \multicolumn{1}{r|}{1}         & 13.12                           & 0.69                            & 1.02                           & 4                               & \multicolumn{1}{l|}{21.49}   & 19.56                       & 1.26                            & 1.01                           & \multicolumn{1}{r|}{11}    & 19.54                       & 1.26                            & 0.99                           & 18                          & \multicolumn{1}{l|}{28.65}   & 24.67                       & 1.33                            & 1.01                           & \multicolumn{1}{l|}{29}    & 24.23                       & 1.33                            & 1.00                           & 21                          \\
			\(50\)                 & \multicolumn{1}{l|}{14.25}              & 13.17                                   & 0.69                                    & 1.03                           & \multicolumn{1}{r|}{1}         & 13.13                           & 0.69                            & 1.07                           & 1                               & \multicolumn{1}{l|}{21.37}   & 19.56                       & 1.30                            & 1.05                           & \multicolumn{1}{r|}{1}     & 19.30                       & 1.30                            & 1.02                           & 4                           & \multicolumn{1}{l|}{28.49}   & 25.17                       & 0.69                            & 1.02                           & \multicolumn{1}{l|}{11}    & 25.28                       & 0.69                            & 0.99                           & 18                          \\ \hline
		\end{tabular}
	}
\end{table}
\begin{table}[ht]
	\centering
	\caption{The best-performing tuple of \((\alpha,\beta,c)\) for the denoising experiment on \textit{SST} and \textit{COVID19-USA} datasets.}\label{tab:denoise:tuple}
	\resizebox{\linewidth}{!}{
		\begin{tabular}{@{}llllllllllll@{}}
			\toprule
			\multirow{2}{*}{\textbf{Dataset}}  & \multirow{2}{*}{\textbf{Method}}               &                                             & \multicolumn{3}{c}{\(k=2\)} & \multicolumn{3}{c}{\(k=5\)} & \multicolumn{3}{c}{\(k=10\)}                                                                                                                                                                         \\ \cmidrule(l){4-6}\cmidrule(l){7-9}\cmidrule(l){10-12}
			                                   &                                                & \multicolumn{1}{c}{\(\boldsymbol{\sigma}\)} & \multicolumn{1}{c}{0.10}    & \multicolumn{1}{c}{0.15}    & \multicolumn{1}{c}{0.20}     & \multicolumn{1}{c}{0.10}  & \multicolumn{1}{c}{0.15}  & \multicolumn{1}{c}{0.20}  & \multicolumn{1}{c}{0.10}  & \multicolumn{1}{c}{0.15}  & \multicolumn{1}{c}{0.20}  \\ \midrule
			\multirow{2}{*}{\textbf{SST}}      & \(\text{\textbf{JFRT}}_{\text{\textbf{Adj}}}\) &                                             & \((1.05, 1.00, 74)\)        & \((1.06, 1.00, 74)\)        & \((0.93, 1.00, 76)\)         & \((1.02, 1.00, 73)\)      & \((0.93, 1.00, 73)\)      & \((0.93, 1.00, 82)\)      & \((0.93, 1.00, 59)\)      & \((1.01, 1.00, 84)\)      & \((0.94, 1.00, 84)\)      \\
			                                   & \(\text{\textbf{JFRT}}_{\text{\textbf{Lap}}}\) &                                             & \((1.05, 1.00, 74)\)        & \((0.94, 1.00, 74)\)        & \((1.07, 1.00, 74)\)         & \((0.93, 1.00, 61)\)      & \((0.94, 1.00, 77)\)      & \((0.94, 1.00, 86)\)      & \((0.99, 1.00, 65)\)      & \((1.02, 1.00, 87)\)      & \((0.94, 1.00, 87)\)      \\ \midrule
			                                   &                                                & \multicolumn{1}{c}{\(\boldsymbol{\sigma}\)} & \multicolumn{1}{c}{0.010}   & \multicolumn{1}{c}{0.015}   & \multicolumn{1}{c}{0.020}    & \multicolumn{1}{c}{0.010} & \multicolumn{1}{c}{0.015} & \multicolumn{1}{c}{0.020} & \multicolumn{1}{c}{0.010} & \multicolumn{1}{c}{0.015} & \multicolumn{1}{c}{0.020} \\ \midrule
			\multirow{2}{*}{\textbf{COVID-19}} & \(\text{\textbf{JFRT}}_{\text{\textbf{Adj}}}\) &                                             & \((0.70, 1.01, 2)\)         & \((0.70, 1.02, 2)\)         & \((0.70, 1.13, 3)\)          & \((0.70, 1.30, 1)\)       & \((0.70, 1.30, 1)\)       & \((0.70, 1.30, 1)\)       & \((0.70, 1.30, 1)\)       & \((0.70, 1.30, 1)\)       & \((0.70, 1.30, 1)\)       \\
			                                   & \(\text{\textbf{JFRT}}_{\text{\textbf{Lap}}}\) &                                             & \((0.70, 1.04, 2)\)         & \((0.70, 1.04, 2)\)         & \((0.70, 1.04, 2)\)          & \((0.70, 0.70, 1)\)       & \((0.70, 0.70, 1)\)       & \((0.70, 0.70, 1)\)       & \((0.70, 0.70, 1)\)       & \((0.70, 0.70, 2)\)       & \((0.70, 0.70, 2)\)       \\
			\bottomrule
		\end{tabular}
	}
\end{table}

\end{document}